\documentstyle[prl,aps,twocolumn,epsf]{revtex}
\def\Vt#1{{ \underline {\hbox{#1}} }}

\def\sxx{\sigma_{xx}}
\def\sxy{\sigma_{xy}}
\def\pbar{\bar p}
\def\pref#1{(\ref{#1})}
\def\eq{\begin{equation}}
\def\eeq{\end{equation}}
\def\eqa{\begin{eqnarray}}
\def\eeqa{\end{eqnarray}}
\def\eff{{\rm eff}}
\def\ext{{\rm ext}}
\def\Sca{{\cal A}}
\def\Scl{{\cal L}}
\def\topic#1{{\medskip \noindent $\bullet$ {\it #1:}}}
\def\pmb#1{\setbox0=\hbox{#1}%
\kern-.025em\copy0\kern-\wd0
\kern.05em\copy0\kern-\wd0
\kern-.025em\raise.0433em\box0}  
\def\bPi{\pmb{$\Pi$}}
\def\bP{{\bf P}}
\def\bV{{\bf V}}

\begin{document}
\twocolumn[\hsize\textwidth\columnwidth\hsize\csname@twocolumnfalse%
\endcsname
\rightline{McGill-00/22, IASSNS-HEP-00/77}
\title{Particle-Vortex Duality and the Modular Group: \\
Applications to the
Quantum Hall Effect and Other 2-D Systems}
\author{C.P.~Burgess$^{1,2}$ and Brian P. Dolan$^{3}$\\
\vspace{3mm}
{\small
$^{1}$ Institute for Advanced Study, Princeton, NJ, 08540.}\\
{\small
$^{2}$ Physics Department, McGill University,
3600 University Street, Montr\'eal, Qu\'ebec, Canada H3A 2T8.}\\
{\small
$^{3}$ Dept. of Mathematical Physics, 
National University of Ireland, Maynooth, Republic of Ireland.}\\
\vspace{3mm}
{\small e-mail: cliff@physics.mcgill.ca, bdolan@thphys.may.ie}}

\date{October 2000}

\maketitle
\begin{abstract}
We show how particle-vortex duality implies the existence of a large 
non-abelian discrete symmetry group which relates the electromagnetic 
response for dual two-dimensional systems in a magnetic field. For conductors 
with charge carriers satisfying Fermi statistics (or those related to fermions
by the action of the group), the resulting group is known to imply many, if 
not all, of the remarkable features of Quantum Hall systems. For conductors 
with boson charge carriers (modulo group transformations) a different group is 
predicted, implying equally striking implications for the conductivities of these 
systems, including a super-universality of the critical exponents for 
conductor/insulator and superconductor/insulator transitions in two dimensions
and a hierarchical structure, analogous to that of the quantum Hall effect
but different in its details.
Our derivation shows how this symmetry emerges at low energies, depending only 
weakly on the details of dynamics of the underlying systems.
\end{abstract}
\bigskip
\hfill{PACS nos: 73.40.Hm, 05.30.Fk, 02.20.-a} 
\bigskip
]

\section{Introduction}
Two-dimensional electron systems have remarkable properties, 
including the quantum Hall effect and metal-insulator transitions, many features
of which still resist theoretical explanation. The difficult part of 
describing these systems is that they involve strong correlations,
and no small parameters present themselves to help with the analysis. This
deprives theorists of most of the tools in their conceptual toolboxes.

Two kinds of theoretical tools which have proven useful for analysing
these kinds of strongly-coupled problems are the exploitation of
symmetries and of the simplifications which are associated with
the low-energy limit. Duality symmetries, in particular, are likely to be
useful since these typically relate strongly-coupled
degrees of freedom to weakly-coupled ones, and in two dimensions
particles and vortices make natural candidates for dual partners. Indeed,
particle-vortex duality has been used several times in the literature
to describe some aspects of both the Quantum Hall effect and conductor-insulator 
transitions in superconducting films. 

Duality symmetries are unusual in that they are not symmetries in the usual
sense that they need commute with the system's Hamiltonian. Instead, they
relate two different kinds of systems to one another.\footnote{Systems for
which dualities commute with the Hamiltonian make up the special case
of self-dual systems.} This relationship is useful when one of the two
systems so related can be analysed, permitting conclusions to be drawn
for its dual by acting with the duality transformation. Alternatively,
duality can be useful if it maps a family of systems into themselves, since
duality invariance then constrains how one flows amongst members of the
family as external parameters -- like temperature or magnetic field -- 
are varied. 

Our aim is to show that both of these lines of argument have very general
applications to two-dimensional systems. In particular, our main result
is to show that the twin operations of particle-vortex 
duality \cite{Fisher,FisherLee,LeeFisher,LeeKane}
and the addition 
of $2 \pi$ units of statistics to charge carriers (which does not change 
their statistics at all) take a very simple form when expressed in terms 
of the material's electromagnetic (EM) response 
functions.\footnote{See references
\cite{WilczekShapere,ReyZee,BCS,DST} for other 
approaches to duality in Quantum Hall
systems.}  These relations
hold for {\it any} system for which the low-energy EM response is dominated by
the motion of quasi-particles or vortices, and for which the dynamics of
these quasi-particles and vortices are similar (in a way we make more explicit
in what follows). Since these two transformations do not commute, they 
generate an infinite discrete group of duality relations amongst the 
EM response functions, and it is this large group which underlies the
predictions which we shall find. 

In general the duality transformations we find do not preserve the
momentum dependence of the EM response function and so, for instance, 
can relate materials whose response is very different (such as by relating
superconductors to insulators). It turns out that conductors are mapped
into themselves, however. When specialized to conductors, with the 
electromagnetic response characterized by the Ohmic and Hall conductivities, 
$\sxx$ and $\sxy$, the action of duality takes the form of subgroups
of the modular group, $PSL(2,Z)$, acting on the complex conductivity, 
$\sigma = \sxy + i \sxx$,\footnote{We use units for which $e^2/h = 1$.} as follows: 
\begin{equation}
\label{sl2zdef}
\tilde\sigma = { a\, \sigma + b \over c \, \sigma + d} ,
\end{equation}
with the integers $a$ through $d$ satisfying $ad - bc = 1$
(the occurrence of this symmetry in a statistical mechanical model
was first noticed in \cite{Cardy}, in an investigation
aimed at understanding QCD).
The duality transformations as defined in eq.~\pref{sl2zdef} are {\it not}
symmetries of the Hamiltonian since, for instance, dual pairs differ in their 
electromagnetic response. Rather, these transformations are symmetries
of the flow obtained as external parameters, such as magnetic fields and
temperatures, are varied. 

Which particular subgroup is important depends on the statistics of the 
charge carrying quasi-particles of the problem. If they are fermions, or
related to them by the symmetries we shall describe, then
the subgroup (denoted $\Gamma_0(2)$ in the mathematics literature \cite{Rankin}) 
is defined by the condition that the parameter $c$ must be even. 
For bose charge carriers, or their symmetry partners,
such as for superconducting films or Josephson Junction
arrays, the symmetry is instead $\Gamma_\theta(2)$, defined by the condition that
$a,d$ are both odd and $b,c$ are both even, or vice versa. For both
cases powerful predictions follow from the fact that our derivation shows
that the symmetry constrains how the conductivities change as
external variables are varied.

For Quantum Hall systems, the group $\Gamma_0(2)$ has 
been long conjectured to be important \cite{LutkenRoss,KLZ,Lutken}, and
has been derived for these systems within a mean-field approximation \cite{KLZ}.
For these systems our derivation accomplishes two new things. First, although
our arguments are modelled on those of ref.~\cite{KLZ}, ours have a broader 
domain of validity since they explicitly assume only that quasi-particles or
vortices dominate the low-energy EM response, and that the systems are clean
enough to exclude any interactions which might distinguish quasi-particles
from vortices, and so thereby ruin the duality symmetry which relates them. Because
the fields we treat only arise in the effective theory, and are not meant to
describe the complete electron dynamics, the mean-field approximation is
kept under better control. 

Second, our derivation helps clarify the assumptions which
underlie analyses of the consequences of $\Gamma_0(2)$ invariance for
the renormalization-group (RG) flow in the $\sxx-\sxy$ plane, since we show
that this only relies on the underlying particle-vortex duality and on the 
long-wavelength limit. This is important because it has been 
shown \cite{Lutken,BD,Semicircle} that most of the unique features 
of Quantum Hall electromagnetic response follow from the consistency
of $\Gamma_0(2)$ invariance with RG flow in the $\sxx-\sxy$ plane, independent
of the detailed form of the flow's $\beta$-function. (The constraints on 
$\beta$ which follow from this symmetry have also been considerably explored
\cite{BD,LutkenBurgess,Crossover,Taniguchi,B+L}.) Previously the key assumption
of two-dimensional flow, governed by $\Gamma_0(2)$ invariance, was just that:
an assumption, although a plausible one motivated by analogy with
the two-dimensional scaling theory of 
disorder \cite{AALR,Khem,Pruiskena,Huckestein}. In particular, since 
the scaling theory strictly only applies near the system's critical points 
the scaling motivation could not explain why many of the predictions 
following from $\Gamma_0(2)$-invariant flow work extremely well, 
even away from the flow's critical points.
Since the derivation presented here is not similarly restricted to 
scaling regions, it explains why these otherwise surprisingly successful 
predictions work.

The identification of $\Gamma_\theta(2)$ as the duality symmetry for 
two-dimensional conductors with bosonic charge carriers is new, although
some precursors of this idea exist. 
Implications of the particle-vortex generator 
of the group for critical behaviour in a superconductor-insulator 
transition have been examined \cite{Fisher}, and the group $\Gamma_\theta(2)$
was discussed as potentially playing a role in two-dimensional systems 
\cite{LutkenRoss,Lutken}, including possibly for the 
Quantum Hall effect \cite{WilczekShapere}. The action of
$\Gamma_\theta(2)$ on longitudinal conductivities 
was written down in \cite{Pryadko}, 
although this was not extended to the whole upper-half
complex plane.

Also new to this paper are the detailed predictions
which follow for bosonic systems from the proposed symmetry group, and which
are direct analogs of the symmetry consequences which are already known
for Quantum Hall systems. 

To which real-life systems does our duality symmetry apply? Just as it is
difficult to compute {\it ab initio} whether a material will be a solid
or not, it is difficult to answer from first principles which systems must be
particle-vortex symmetric. There are several things which can be said, however.
First, if the symmetry holds, then all of its consequences follow together.
For example, for Quantum Hall systems the semi-circle law, $\rho_{xx} \to
1/\rho_{xx}$ duality, super-universality of the critical exponents,
universality of the transition conductivities, odd-denominator 
fractional quantization of
$\sxy$ when $\sxx = 0$, the selection rules for which plateaux may
be related by transitions, {\it etc.} must all come together as a package. 

Second, since particle-vortex duality relies on the equivalence of
the kinematics and couplings of the charge-carrying quasi-particles and
vortices, it should be a good approximation when the only quantities of
interest in the Hamiltonian are those describing the kinematics of these
particles, and their couplings to the fields which describe the long-range
vortex interactions and the electromagnetic fields which are
applied to test the EM response. 
Duality could be ruined by other microscopic interactions which treat
quasi-particles and vortices differently, such as from couplings 
with disorder or with other electronic degrees of freedom. Of course,
disorder can also play other spoiling roles, such as by destroying the
phase coherence on which the quantum regime which we assume depends.

A sufficient condition for particle-vortex duality, and its 
associated non-abelian extensions, might therefore be that the system
be sufficiently clean to justify the neglect of other particle and vortex interactions 
when calculating the electromagnetic response. Although this condition is
not strictly necessary, since our derivation also applies in the presence of
any particle/vortex-democratic interactions, weak coupling is also implicit
in our neglect of anomalous dimensions when deciding the relevance or irrelevance
of low-energy interactions. Notice that it is the weakness of the quasi-particle
couplings which are important in this decision, and the assumption that these
are weak does not imply that the underlying electrons must also be weakly coupled
in the microscopic Hamiltonian.

Now to the main arguments. 
We organize our presentation as follows: First we describe the action
describing the low-energy dynamics of a system of quasi-particles and
vortices, and cast it into a form which emphasizes the similarities
between these two kinds of charge carriers. Next, we derive the action
of the two basic symmetries --- $2\pi$ statistics addition and particle-vortex
duality --- for the electromagnetic response functions. Then, we specialize
the result to the particularly interesting case of a conductor, to derive
the action of the symmetry on the conductivities, $\sxx$ and $\sxy$. 
We briefly review the Quantum Hall case, where the charge carriers are
fermions, and then repeat the analysis for charge carriers satisfying
bose statistics, listing many experimental predictions which follow
from the symmetries. 

\section{Particles and Vortices}
This section has two goals. First, we derive an expression for the effective action 
governing the low-energy interactions of charged quasi-particles and vortices for
which the duality between these two kinds of objects is made explicit. Second,
we compute the EM response for a system of such charges and vortices, for use
when deriving the implications of particle-vortex duality in the next section.

\subsection{The Effective Quasi-particle-Vortex Action}
Our starting
point is the following lagrangian, which describes the low-energy/long-wavelength
interactions of a collection of $N_p$ quasi-particles and $N_v$ vortices 
with a weak electromagnetic probe, $A_\mu$:
\eqa
\label{LELWLagr}
\Scl_\eff &=& - \, {\pi \over 2 \theta} \; \epsilon^{\mu\lambda\nu} \,
a_\mu \partial_\lambda a_\nu + \Scl_p(\xi,a+A) \nonumber\\
&& - \frac{\kappa}{2} \, [\partial_\mu \phi
- q_\phi (a_\mu + A_\mu)][\partial^\mu \phi - q_\phi(a^\mu + A^\mu)] \nonumber\\
&& \qquad\qquad + \cdots ,
\eeqa
which we write in the continuum approximation because our interest
is directed towards the low-frequency, long-wavelength EM response. 

Several features of this lagrangian bear explanation, and since our 
final results ultimately depend on its validity, we pause here to explain 
its form in some detail. 

\topic{Why Both Particles and Vortices?}
Usually charged particles and vortices, in the way we define them, do not coexist
in the low-energy theory, since the vortices presuppose the breaking of 
electromagnetic symmetries which precludes the existence of isolated electrically 
charged particles, and our later applications only require the consideration of
systems containing one or the other. We nonetheless use the mathematical device of
keeping both in eq.~\pref{LELWLagr} since it permits us to derive our results for
particles only and vortices only by taking the appropriate limits of a single formula. 

\topic{Field Content}
$A_\mu$ is the small external electromagnetic potential which 
is applied to probe the system's EM response, $\xi^\mu_k(t)$ is the position
of the $k$'th quasi-particle of the system as a function of time, 
and $a_\mu$ is the usual statistical gauge potential which ensures that the 
interchange of two quasi-particles produces the phase $e^{i\theta}$ \cite{ASW}. 
($\theta = 2 \pi n$, for integer $n$, corresponds to bosonic quasi-particles, 
while $\theta = (2n +1)\pi$ describes fermionic quasi-particles.)\footnote{Appendix
A briefly reviews our conventions concerning this statistics field.}

\topic{Relativistic Form}
We use relativistic notation in eq.~\pref{LELWLagr} in order to most cleanly 
illustrate the logic of the argument. This proves to be convenient because
the relativistic case shows all of the main features of duality, and is 
considerably simpler to describe. We have checked that other features of
a non-relativistic treatment -- such as the potential appearance of kinetic terms 
linear in time derivatives -- do not substantially affect our arguments.
It also happens that the results for the non-relativistic systems of practical 
interest can be read off directly from the relativistic answers using the trick 
outlined in Appendix B. 

\topic{Quasi-particle Lagrangian} 
$\Scl_p(\xi,a+A)$ is the (first-quantized) lagrangian which
describes the quasi-particle motion and their coupling to the
electromagnetic field, $A_\mu$. A first-quantized representation
is chosen because this makes the duality between particles and vortices
most transparent later on. The detailed form of this lagrangian is not
important in what follows, apart from the form of the coupling
to the gauge field, $(a+A)_\mu$, whose form is important, but which
follows on grounds of gauge invariance. 
To be concrete, in the absence of other interactions
the particle lagrangian might be given explicitly by
\eq
\label{pact}
\Scl_p = \sum_k \left[\frac{m}{2} \dot{\xi}^\mu_k \dot{\xi}_{k\mu} 
-q_k \dot{\xi}^\mu_k (a+A)_\mu \right] \delta[x-\xi_k(t)], 
\eeq
where $q_k$ here denotes the quasi-particle charge,
normalized so that $q_k = -1$ for electrons. 

\topic{Dependence on External Parameters}
All of the dependence on the external variables,
like magnetic field, $B$, enter through the parameters
of the effective lagrangian. For instance, in the example where the quasi-particle 
dynamics is described by eq.~\pref{pact} they would
enter through the quasi-particle mass, $m$, and the parameter $\kappa$,
although the dependence would be more involved for more complicated
quasi-particle dynamics. The same is almost true for the dependence on
temperature, $T$. That is, the temperature dependence contributed when
integrating out high-energy modes is embedded in the system parameters,
but there is also additional temperature dependence associated with the
integration over the low-energy degrees of freedom themselves, such as
$\xi^\mu_k$. Both forms of temperature dependence are included in our later
discussions of the temperature-dependence of particle-vortex duality.

\topic{The Goldstone Variable} 
If the difference between the number of vortices and anti-vortices, $N_v$, is
nonzero then there is a complex order parameter somewhere
in the system which typically vanishes at the positions of the vortices (and
anti-vortices), and takes a nonzero value asymptotically far away. The number $N_v$ 
is then related to the winding of the phase of this order parameter
around a circle which encloses all of the vortex positions. The field
$\phi$ in eq.~\pref{LELWLagr} represents the phase of this order parameter. 
We assume, in writing eq.~\pref{LELWLagr}, that the order parameter
carries nonzero electric charge, $q_\phi \ne 0$, and so it spontaneously 
breaks the electromagnetic $U(1)$ gauge group. The scale of the parameter $\kappa$
is of order of the scale of the symmetry-breaking expectation value.

Since the quanta of $\phi$ are the Goldstone bosons for the assumed symmetry
breaking, $\phi$ is guaranteed to be in the low-energy theory. Indeed $\phi$
would be responsible for the long-range interactions experienced between 
vortices, if $q_\phi$ were zero. Furthermore,
its couplings to $(a+A)_\mu$ are dictated by gauge invariance. Finally, because
it is a Goldstone variable $\phi$ is guaranteed by Goldstone's theorem to
decouple at low energies (modulo the usual
Coleman-Mermin-Wagner caveats), thereby justifying its semi-classical treatment
using a derivative expansion \cite{GBreview}. The ellipses in eq.~\pref{LELWLagr}
represent all of the other effective interactions obtained when all higher
degrees of freedom are integrated out. Since these all involve inverse powers
of the higher-energy scales, such as the order parameter scale,
they are irrelevant for the present purposes compared to those 
explicitly displayed.

In the derivation of KLZ \cite{KLZ}
this order parameter was the bosonic field which described the electrons, but
for the present purposes it could equally well be a bilinear of fermion fields,
or something still more complicated. All we need assume is that the order
parameter spontaneously breaks electromagnetic gauge invariance, and that its
boundary conditions at spatial infinity incorporate the winding corresponding
to vortex number $N_v$. 

\subsection{A Dual Description of the Vortices}
In order to better display the particle-vortex duality, it is convenient
to use the dual description of the vortex degrees of freedom 
\cite{vortexdual,LeeZhang,Zhang}.
We here implement this duality transformation by recognizing that
it is a special case of a general
dualization algorithm \cite{buscher} (which also has applications
to bosonization in one and two dimensions \cite{bosonization}), and
is derived in detail for the present system in Appendix C. (We repeat the 
derivation in a second-quantized format in Appendix D, using the same 
model -- the nonrelativistic abelian Higgs model supplemented with Chern Simons
coupings -- used by KLZ.)

The result, when applied to the quasi-particle/vortex action, eq.~\pref{LELWLagr},
is:
\eqa
\label{dualLagr}
\tilde\Scl_\eff &=& - \, {\pi \over 2 \theta} \; \epsilon^{\mu\lambda\nu} \,
a_\mu \partial_\lambda a_\nu - \epsilon^{\mu\lambda\nu} \,
b_\mu \partial_\lambda (a_\nu + A_\nu) \nonumber\\
&& + \Scl_p(\xi,a+A) + \Scl_v(y,b) 
  + \dots ,
\eeqa
where:

\topic{The field $b_\mu$} is the new (gauge potential) field which is the dual
representation of the Goldstone field, $\phi$. It carries all
of the information about the long-range interactions amongst the 
vortices. 

\topic{The variables $y^\mu_a(t)$} label the positions of the
centres of the vortices and anti-vortices defined, for instance,
as the positions of the zeros of the underlying order parameter
whose phase was $\phi$. As is seen in Appendix C, these positions naturally
arise as variables during the duality transformation once one
takes into account the boundary conditions satisfied by $\phi$ 
in the presence of vortices.

\topic{The Vortex Lagrangian ${\cal L}_v(y,b)$} describes the dynamics of
the vortices and their couplings to the field $b_\mu$. Although
this action can be complicated, reflecting the potentially
complicated dynamics of vortices in the material of interest, it
must have at least the following two terms:
\eq
\label{vact}
\Scl_v = \sum_a \left[\frac{M}{2} \dot{y}^\mu_a \dot{y}_{a\mu} 
-\,{2\pi  N_a \over q_\phi} \; \dot{y}^\mu_a b_\mu \right] \delta[x-y_a(t)], 
\eeq
where $N_a$ denotes the vorticity (or winding number) of each vortex.
The coupling term between $y^\mu_a$ and $b_\mu$ falls directly out
of the duality transformation, and so is quite generally known. 
It should be noticed that although the kinetic term for $y^\mu_a$ does
not itself follow directly by dualizing eq.~\pref{LELWLagr}, its form
is fixed quite generally from symmetry arguments \cite{Nambu}. This is
because the $y^\mu_a$'s may themselves be thought of as Goldstone
bosons for the breaking of translation invariance, which is here broken by
the positions of the vortices themselves. In general there may
also be other interactions to supplement eq.~\pref{vact}, which
describe the interactions of the vortices with other degrees of
freedom, such as disorder. 

\topic{A $b_\mu$ kinetic term} of the form  
$- \frac{1}{4\kappa q_\phi^2} \, f^b_{\mu\nu} (f^b)^{\mu\nu}$, with
 $f^b_{\mu\nu} = \partial_\mu b_\nu - \partial_\nu b_\mu$,
  is also produced
when performing the dualization, but is not written in eq.~\pref{dualLagr}.
It has been dropped since it is inversely proportional to $\kappa$, which is
one of the high-energy scales whose inverse we are ignoring in the
low-energy, long-wavelength limit. It would compete with a term proportional to
$(f_{\mu\nu} + F_{\mu\nu})(f^{\mu\nu} + F^{\mu\nu})$ (where $f_{\mu\nu}$
and $F_{\mu\nu}$ are the field strengths for $a_\mu$ and $A_\mu$), which
was among the ellipses appearing in eq.~\pref{LELWLagr}. (Alternatively, both
such terms may be absorbed into the general response function which is obtained
when $\xi^\mu_k$ and $y^\mu_a$ are integrated out, as we are shortly to describe.)

\topic{Equivalence} It is to be emphasized that eq.~\pref{dualLagr} is
just a change of variables of eq.~\pref{LELWLagr}, and so describes
precisely the same physics. In particular, both expressions reproduce
precisely the same electromagnetic response once all degrees of freedom
except for $A_\mu$ are integrated out:\footnote{Although we present our results for
real time and at zero temperature, our duality relations can be derived
equally well in imaginary time at nonzero temperature.}
\eqa
\label{EMRdef}
e^{i\Gamma(A)} &=& \int [da_\mu(x)] \, \prod_k [d\xi^\mu_k(t)] \;
\exp\left[ i \int d^3x \Scl_\eff(\xi,a,A) \right] \nonumber\\
&=& \int [da_\mu(x)][db_\mu(x)] \, \prod_a [dy^\mu_a(t)]\nonumber\\
&&\qquad\qquad \times\exp\left[ i \int d^3x \tilde\Scl_\eff(y,a,b,A) \right] .\nonumber\\
&&
\eeqa

\section{Particle-Vortex Duality}
The beauty of eqs.~\pref{pact}, \pref{dualLagr} and \pref{vact} is
that they display the quasi-particle and vortex degrees of freedom in
a way which emphasizes the similarity of the particles and vortices.
For instance, if other interactions are negligible, so the particle and 
vortex dynamics is given by eqs.~\pref{pact} and \pref{vact}, and
in the absence of the Chern-Simons term involving $\theta$, eq.~\pref{dualLagr}
has the symmetry with respect to the interchanges
$\xi^\mu_k \leftrightarrow y^\mu_a$, $b_\mu \leftrightarrow
a_\mu + A_\mu$, so long as the quasi-particle masses and charges are also
interchanged: $q_k \leftrightarrow 2\pi N_a/q_\phi$ and $m \leftrightarrow M$. 
The same is also true if $\Scl_p$ and $\Scl_v$ are more complicated,
provided that the additional complications treat particles and 
vortices democratically, by contributing terms of the same form to
both $\Scl_p$ and $\Scl_v$. 

We may now be more precise as to what is meant by dual systems. Given
a two-dimensional system whose EM response is governed by
$n$ quasi-particles (having mass $m$ {\it etc}) and $N$ vortices (having
mass $M$ {\it etc}), we define the dual to be the system having
$N$ quasi-particles (with mass $M$ {\it etc}) and $n$ vortices 
(of mass $m$ {\it etc}). The cases of real interest for the applications
which follow are the cases involving particles only or vortices only: 
$n=0$ and $N=0$.

\subsection{Expressions for the EM Response}
We now wish to determine how the electromagnetic response of dual
systems are related to one another. To do so, imagine integrating out
the quasi-particles and vortices by performing the path integral over
their positions, $\xi_k^\mu$ and $y^\mu_a$. If we are only interested
in linear response, we need not actually evaluate these integrals, but
may parameterize them in terms of response functions, $P^{\mu\nu}$,
and $V^{\mu\nu}$, as follows. That is, if we define:
\eqa
\label{RFdef1a}
e^{i\gamma_p(a+A)} &=& \int \prod_k [d\xi^\mu_k(t)] \;
\exp\left[ i \int d^3x \Scl_p(\xi,a+A) \right] \\
\label{RFdef1b}
e^{i\gamma_v(b)} &=& \int \prod_a [dy^\mu_a(t)] \;
\exp\left[ i \int d^3x \Scl_v(y,b) \right] ,
\eeqa
then for linear response it suffices to take:
\eqa
\label{RFdef2a}
\gamma_p(a) &=& -\, \frac12 \int d^3x\, d^3x' \; a_\mu(x) P^{\mu\nu}(x-x') a_\nu(x'),\\
\label{RFdef2b}
\gamma_v(b) &=& -\, \frac12 \int d^3x\, d^3x' \; b_\mu(x) V^{\mu\nu}(x-x') b_\nu(x') ,
\eeqa
where $P^{\mu\nu}(x-x')$ and $V^{\mu\nu}(x-x')$ define the particle and
vortex response functions. Notice that for dual systems we expect $P^{\mu\nu}
= \tilde{V}^{\mu\nu}$ and $V^{\mu\nu} = \tilde{P}^{\mu\nu}$, where the 
tilde denotes the result evaluated in the dual system. 

If eqs.~\pref{RFdef1a} -- \pref{RFdef2b} are used in (\ref{dualLagr}), 
then the remaining 
integrals over $a_\mu$ and $b_\mu$ are Gaussian and so may be evaluated
explicitly to obtain the electromagnetic response function, defined by:
\eq
\label{PIdef}
\Gamma(A) = -\, \frac12 \; \int d^3x \, d^3x' \; A_\mu(x) \Pi^{\mu\nu}(x-x') A_\nu(x').
\eeq
One might be queasy about the consistency of first expanding to quadratic order
in $b_\mu$ and then integrating $b_\mu$ over all values, and this queasiness would be justified 
if the expansion to quadratic order was done because $b_\mu$ is small. Such a calculation 
really presumes an effective-field theory approach, where all terms involving higher powers 
of fields are irrelevant (in the RG sense), and so may be neglected within the very low energy 
effective theory. This is true in the present instance so long as anomalous dimensions
are small, so that relevance may be judged purely using na{\rm\"\i}ve dimensional analysis. It is
here that we implicitly assume the quasi-particles and vortices to be weakly coupled to
other degrees of freedom, although this does not also imply weak coupling for the underlying
electrons. In any case, we later provide an alternative derivation of a subset of our results 
which does not rely on the quadratic approximation of eqs.~\pref{RFdef2a}, \pref{RFdef2b}.

We believe that our treatment of the quasi-particles purely within an 
effective-theory framework represents 
an important conceptual difference between the arguments formulated
here, and those presented in the spirit of KLZ, in Appendix D. They differ because in the 
KLZ framework the bosonic field whose phase describes the vortices is taken to be a
direct description of the underlying electrons, rather than a low-energy effective field. 
As a result a weak-coupling mean-field analysis for the KLZ field is more directly tied to
the strength of the couplings of the underlying electrons.

Before quoting the result obtained in this way for the EM response, 
it is worth first defining some
notation. The polarization tensor is usually taken to have the standard
rotationally-invariant and gauge-invariant but non-relativistic form:
\eqa
\label{nrccform}
\Gamma(A) &=& \frac12 \int d^3x \, d^3x' \; \Bigl[ \pi_1(x-x') E_i(x) E_i(x')
\nonumber\\
&&\qquad - \pi_2(x-x')  B(x) B(x') \\
&&\qquad - \pi_3(x-x') \epsilon^{\mu\nu\lambda} A_\mu(x)
\partial_\nu A_\lambda(x'), \nonumber
\eeqa
defining the electromagnetic form factors, $\pi_1, \pi_2$ and $\pi_3$. In what
follows our main interest is in $\pi_1$ and $\pi_3$, which control the 
conductivities $\sxx$ and $\sxy$. Because of this, and because of the greater
simplicity of the resulting formulae, we specialize instead to the
relativistic version of eq.~\pref{nrccform}, which we write in momentum
space as:
\eq
\label{FFactdef}
\Pi^{\mu\nu} = \Pi_1(p^2) \; \Lambda^{\mu\nu} + \Pi_3(p^2) \; J^{\mu\nu},
\eeq
where ${\Lambda^\mu}_\nu$ and $J^{\mu\nu}$ are defined by
${\Lambda^\mu}_\nu = \delta^\mu_\nu - p^\mu p_\nu/p^2$ and $J^{\mu\nu} =
i\epsilon^{\mu\lambda\nu} \, p_\lambda/\sqrt{p^2}$. Because
the form factors of eqs.~\pref{nrccform} and \pref{FFactdef} are related 
(in momentum space) by $\Pi_1 = (p/\hbar)^2 \,\pi_1 = (p/\hbar)^2 \pi_2$ and 
$\Pi_3 = \sqrt{p^2/\hbar^2} \; \pi_3$,
the relativistic form is sufficient to follow how the quantities $\pi_1$ and 
$\pi_3$ transform under duality transformations. (More generally, results
for the generic case $\pi_1 \ne \pi_2$, are easily obtained using the trick described
in Appendix B.)

The great utility of the relativistic expression, eq.~\pref{FFactdef}, follows
because the tensors ${\Lambda^\mu}_\nu$ and $J^{\mu\nu}$ satisfy the identities:
${\Lambda^\mu}_\alpha \, {\Lambda^\alpha}_\nu = {\Lambda^\mu}_\nu$, ${\Lambda^\mu}_\alpha
J^{\alpha\nu} = J^{\mu\alpha} {\Lambda^\nu}_\alpha = J^{\mu\nu}$ and
$J^{\mu\alpha} J_{\alpha\nu} = -  {\Lambda^\mu}_\nu$. Since the tensors
$\Lambda$ and $J$ are related to one another in the same way as are the 
bases, 1 and $i$, of complex numbers, tensor manipulations 
with $\Pi^{\mu\nu}$ can be greatly
simplified by re-expressing it as a complex variable: 
\eq
\label{cvdefs}
 \bPi = \Pi_1 + i \Pi_3.
\eeq

With this notation, and defining similar expressions for the complex quantities
$\bP$ and $\bV$ in terms of the corresponding form factors $P_1$, $P_3$, $V_1$
and $V_3$, we find the result of integrating the fields $a_\mu$ and $b_\mu$
out of eq.~\pref{dualLagr} to be
\eq
\label{PiPV}
\bPi = i\sqrt{\pbar^2} \; \left({\pi \over \theta} \right) {\pbar^2 +
\bV \, \bP \over \pbar^2 + \bV \left(\bP + i \sqrt{\pbar^2} \; \left( {\pi \over
\theta} \right)\right) } 
\eeq
as the general relation between the EM response and the particle and
vortex response functions. To avoid explicit factors of $\hbar$
we have defined $\pbar=p/\hbar$ in this expression. 
(The details of these integrations are given
in Appendix E.) This expression is the main result on which
our later conclusions are based. 

We now record the special cases of this expression which will be used
in the following sections.

\topic{Addition of $2\pi$ Flux} Since $\theta$ represents the statistics
of the quasi-particles, the choice $\theta = 2 k \pi$, with $k$ an integer, 
can have no physical effects: $\bPi(\theta + 2k\pi) \approx \bPi(\theta)$. 
Inspection of eq.~\pref{PiPV} shows the implications of this statement 
for the EM response:
\eq
\label{2pishift}
{1 \over \bPi(\theta)} \approx 
{1 \over \bPi(\theta + 2k\pi)} = {1\over \bPi(\theta)} - {2ki \over \sqrt{\pbar^2}},
\eeq
which reproduces a well-known result \cite{KLZ,Fradkin}

\topic{Quasi-particles Only}
Should there be no vortices participating in the EM response at all 
(which, because of the duality transformation from $\phi$ to $b_\mu$,
is equivalent to ${\bf V}\rightarrow\infty$), 
then eq.~\pref{PiPV} simplifies to:
\eq
\label{PiPVp}
\bPi = i\sqrt{\pbar^2} \; \left({\pi \over \theta} \right) {\bP \over 
\bP + i \sqrt{\pbar^2} \; \left( {\pi \over \theta} \right) } .
\eeq
which reduces to $\bPi = \bP$ when $\theta \to 0$, corresponding to
bose statistics for the charge-carrying quasi-particles. The corresponding
result for fermionic charge carriers is similarly found by choosing
$\theta = \pm\pi$. 

\topic{Vortices Only}
In the event that only vortices are involved in charge transport, 
expression \pref{PiPV} reduces to:
\eq
\label{PiPVv}
\bPi =  {\pbar^2 \over \bV - i \sqrt{\pbar^2} \; \left( {\theta\over
\pi} \right)} .
\eeq
This also reduces to the usual result, $\bPi = \pbar^2/\bV$, for bosonic
charge carriers, for which $\theta \to 0$.

\section{Some Consequences of Particle-Vortex Duality}
In this section we derive the implications of particle-vortex duality
for the electromagnetic response of two-dimensional systems. 

We are interested in the situation where quasi-particles only or 
vortices only are responsible for charge transport, in which case a 
very simple expression may be derived for the 
action of particle-vortex duality on the EM response function, $\bPi$.
This may be derived by using eq.~\pref{PiPVp} to relate $\bPi$ to $\bP$
for the original system, and using eq.~\pref{PiPVv} to relate $\tilde{\bPi}$
to $\tilde{\bV} = \bP$ for the dual system. Eliminating $\bP$ from these two
expressions gives the desired relation
\eq
\label{PitwvsPi}
{\tilde{\bPi} \over \pbar^2} =  {i \sqrt{\pbar^2}\; \left({\pi\over \theta}\right)
 - \bPi \over \pbar^2 + i \sqrt{\pbar^2} \left( {\pi\over \theta } + {\theta \over \pi}
 \right) \; \bPi} 
\eeq
directly expressing the dual response function, $\tilde{\bPi}$, in terms
of that of the original system. 

The physical interpretation of expression \pref{PitwvsPi} is obtained by using
the connection between the small-$p^2$ limit of the form factors,
$\Pi_1$ and $\Pi_3$, and measurable electromagnetic quantities. There
are two main cases to consider:

\subsection{Superconductors and Insulators}
If there there is a gap at the relevant part of the spectrum, then 
one expects the EM response function to be analytic in momentum space.
This implies the functions $\Pi_1$ and $\Pi_3$ have the following
small-$p^2$ form;
\eq
\label{gapform}
\Pi_1 = A_1  + B_1 \, \pbar^2 + \dots, \qquad \Pi_3 = \sqrt{\pbar^2} \, ( A_3 + \dots).
\eeq
The $\sqrt{\pbar^2}$ does not conflict with $\Pi^{\mu\nu}$ being analytic
as $p^\mu\to 0$ because it cancels a similar factor in the definition of
the tensor $J^{\mu\nu}$. The constant $A_1$ is nonzero only for 
superconductors, with $A_1$ inversely related to the medium's
electromagnetic penetration depth. If $A_1=0$ then the material is an 
insulator and $B_1$ is related to the dielectric response of the medium.
The constant $A_3$ corresponds to a Hall conductivity for the
system, which may vanish but need not. 

There are two important sub-cases to consider:

\subsubsection{Superconductors}
If $A_1$ is nonzero, then the material is a 
superconductor and $A_1 = m_\gamma^2 = 1/\lambda$ is the 
corresponding photon squared mass, 
or inverse penetration length. In this case there is a gap
because of the Anderson-Higgs mechanism. 

Inserting assumption \pref{gapform} for $\bPi$ (with $A_1\ne 0$) into
the duality expression, eq.~\pref{PitwvsPi}, implies $\tilde{\bPi}$ also
has an expansion of the form \pref{gapform}, with:
\eqa
\label{dualscgapform}
\tilde A_1 &=& 0, \nonumber \\
\tilde A_3 &=& {1 \over {\theta\over \pi} + {\pi \over \theta}} ,\\
\tilde B_1 &=& {1\over \left[ 1
+ \left({\theta \over \pi}\right)^2\right]^2 \, A_1 }.
\nonumber
\eeqa
We see that the system dual to a superconductor is an insulator, whose
Hall conductivity and dielectric function are related to the superconducting
penetration length and the statistics of its quasi-particle charge carriers.

\topic{Bosonic Charge Carriers}
A particularly interesting special instance of expression \pref{dualscgapform}
is the case of bosonic charge carriers (such as Cooper pairs), for which
$\theta = 0$. In this case we find as dual a dielectric with vanishing 
Hall conductivity and dielectric function given by $\tilde B_1 = 1/A_1$.
This is as expected on physical grounds  \cite{Fisher} since the condensation
of vortices in the dual system should produce an insulator. Notice that
eq.~\pref{dualscgapform} makes specific predictions as to the dependence
of the dual dielectric function, $\tilde B_1$, as a function of temperature 
since it is related to the temperature dependence of the penetration
length in the superconducting system. 

\topic{Fermionic Charge Carriers}
If we instead choose $\theta = \pi$, as is appropriate for
fermion charge carriers, we find the dual system has no Ohmic
resistance, but has Hall conductivity $\sxy = \frac12$. 

\subsubsection{Insulators}
If $A_1=0$ then the material is an insulator and the gap exists because
of the absence of low-energy charged quasi-particles which carry current.
In this case $B_1$ is related to the dielectric response of the medium
and the duality transformation produces the following small-$p^2$
EM response:
\eqa
\label{dualinsgapform}
\tilde A_1 &=& 0, \nonumber \\
\tilde A_3 &=& {{\pi\over \theta} - A_3 \over 1 - \left( {\theta\over \pi} 
+ {\pi \over \theta} \right) A_3} ,\\
\tilde B_1 &=& {B_1 \left( {\pi\over \theta} \right)^2 \over 
\left[ 1 - \left( {\theta\over \pi} 
+ {\pi \over \theta} \right) A_3 \right]^2 } .
\nonumber
\eeqa
We see that the image of an insulator is another insulator, although
with different Hall conductivity and dielectric function.  

\subsection{Conductors}
For conductors the form factor $\Pi_1$ is not analytic as $p^2 \to 0$.
The limiting form for small $p^2$ is related to the conductivities by:
\eqa
\label{conddef}
\Pi_1(\pbar^2) &\to& \sxx \; \sqrt{\pbar^2} + \dots,\\
\Pi_3(\pbar^2) &\to& \sxy \; \sqrt{\pbar^2} + \dots , \nonumber
\eeqa
and so the complex quantity $\bPi$ is related to the complex
conductivity, $\sigma = \sxy + i \sxx$, by $\bPi = i \sqrt{\pbar^2} \;\sigma^*$.

In this case the duality transformation, eq.~\pref{PitwvsPi}, preserves
the momentum dependence of the form factors, implying $\tilde{\bPi}
= i \sqrt{\pbar^2} \; \tilde\sigma^*$, with the dual conductivities given by
the holomorphic relation:
\eq
\label{dualcond}
\tilde\sigma = {{\pi \over \theta} - \sigma \over 1 - \left( 
{\theta \over \pi } + {\pi \over \theta} \right) \, \sigma}.
\eeq

\topic{Fermion Charge Carriers}
If we assume the original charge carriers to be fermions, as would
be appropriate for the integer Quantum Hall systems, then 
eq.~\pref{dualcond} reduces to the expression:
\eq
\label{dualcondf}
\tilde\sigma = {1 - \sigma \over 1 - 2 \, \sigma},
\eeq
which is the transformation which has been recognized \cite{Semicircle}
to imply the experimentally observed
duality transformations $\rho_{xx} \to 1/\rho_{xx}$
in the transitions between the $\sxy = 1$ 
Hall plateaux and the Hall insulator. 

Duality relations between more complicated Hall plateaux may be
similarly obtained by starting from the appropriate anyonic
charge carriers, but the same result may be obtained more simply by
combining eq.~\pref{dualcondf} with transformation \pref{2pishift} (which
expresses the the absence of content of a shift of the statistics 
parameter by $2\pi$) specialized
to conductors:
\eq
\label{2pishiftcond}
\sigma \approx {\sigma \over 1 - 2\sigma}.
\eeq
Transformations \pref{dualcondf} and \pref{2pishiftcond}
together generate the group, $\Gamma_0(2)$, which is known
to relate all of the allowed (odd-denominator) Hall states to
one another, and so produces the anyonic Quantum Hall duality relations from
the simpler fermionic one we have just considered. 

\topic{Boson Charge Carriers}
Specializing eq.~\pref{dualcond} to the case $\theta = 0$ also gives 
a simple relation between the dual conductivities:
\eq
\label{dualcondb}
\tilde\sigma = - \; {1\over \sigma} .
\eeq
This transformation, together with eq.~\pref{2pishiftcond}, also
generates an infinite group, denoted $\Gamma_\theta(2)$, which relates
dual conductors with bosonic charge carriers (such as superconducting
films or Josephson junction arrays). 

\subsection{Duality and Flow}

To this point we have found how the EM response of dual systems 
are related to one another. We wish now to understand how the duality
transformations change as external parameters like magnetic field and temperature
are continuously varied. 
The question of how systems change as external variables are varied
is particularly sharp for conductors, since in this case both the
original system and its dual are of the same type ({\it i.e.} they
are both conductors). If one imagines a system tracing out a curve
in the conductivity plane as $T$ and $B$ are varied, the position of
the dual system traces out another curve in the same plane. We wish
to argue that the resulting flow commutes with the action of the
two duality transformations, eqs.~\pref{dualcondf} and \pref{2pishiftcond}
(or \pref{2pishiftcond} and \pref{dualcondb}),
in the conductivity plane. 

The magnetic field strength, and other other microscopic 
properties, enter into the above arguments only by changing the values 
which are taken by the masses and other parameters appearing in the effective
lagrangian, eqs.~\pref{LELWLagr} and \pref{dualLagr}. Temperature partly appears in
the same way, but also appears in the integration over the low-energy degrees
of freedom (such as by rotating to euclidean signature and imposing periodicity
in imaginary time). Now comes the main point. All of the consequences of
duality follow from the statements $P^{\mu\nu} = \tilde{V}^{\mu\nu}$ and 
$V^{\mu\nu} = \tilde{P}^{\mu\nu}$, which relate the response functions of
dual systems (or the analogous statement expressing the addition of
$2\pi$ statistics flux). And these statements are true for all values of 
magnetic field, temperature {\it etc.}, so long as these quantities
do not introduces differences in form between the particle and vortex
lagrangians, ${\cal L}_p$ and ${\cal L}_v$. 

Suppose we now consider a system of quasiparticles, $S(T_0,B_0)$, chosen
for some specific temperature, $T_0$, and magnetic field, $B_0$, involving 
$N$ particles of mass $m$ and charge $q$. Next, suppose that $\tilde B_0$ 
and $\tilde T_0$ are chosen to produce the dual system, $\tilde{S}$, having $N$ 
vortices, also with mass $m$ and vortex charge $q$. That is: $S(\tilde T_0,\tilde B_0) = 
\tilde{S}(T_0,B_0)$. Notice that the existence of such a $\tilde T_0$ and
$\tilde B_0$ is plausible given that it involves the solution of two equations
for two unknowns. For instance, for particle-vortex interchange the two equations
are: $M(\tilde T_0,\tilde B_0) = m(T_0,B_0)$ and $Q(\tilde T_0,
\tilde B_0) = q(T_0,B_0)$, where $M$ and $Q$ are the functions of $T$ and $B$
which define the vortex mass and charge.

Now imagine changing the magnetic field and/or temperature, say to $B' = B_0 (1+ 
\delta_B)$ and $T' = T_0(1 + \delta_T)$. The question we ask is this: if we change
the dual system by the same amount, to $\tilde B' = \tilde B_0 (1 + 
\delta_B)$ and $\tilde T' = \tilde T_0 (1+ \delta_T)$, is the resulting system
still dual to the first? That is, is $S(\tilde T',\tilde B') = \tilde{S}(T',B')$?
The answer is `yes', because the question asked of both systems is the same: what is
the change in response of a system of $N$ objects of mass $m$ and charge $q$ 
as $T$ and $B$ are varied? The only difference between the system and its dual is 
that for $S$ the $N$ objects are particles and for $\tilde{S}$ they are vortices. 

Arguing in the same way for the attachment of $2\pi$ statistics flux, we
see that the entire duality group must commute with the flow through
the conductivity plane as $B$ and $T$ are varied. For fermions this implies
a $\Gamma_0(2)$-invariant flow, while for bosons $\Gamma_\theta(2)$-invariant 
flow is implied.

\subsection{Beyond Linear Response}
In this section we present a slightly different version 
of part of the previous
section's derivation, whose aim is to express the action of particle-vortex
duality on the electromagnetic response without relying on the linear-response
approximation, eqs.~\pref{RFdef2a}, \pref{RFdef2b}. The argument we present
assumes fermion charge carriers, since it relies on using a statistics
parameter $\theta = \pm \pi$. Besides clarifying our derivation, 
we present this separate line of argument because we believe it will
ultimately prove fruitful in its own right by explaining
the experimental evidence \cite{dualSTSCSS} for current-voltage duality 
seen in Quantum Hall systems beyond the linear Ohm's Law approximation.

We pick up the story of section (III) just before making the linear-response
approximations, eqs.~\pref{RFdef2a}, \pref{RFdef2b}. The response function,
$\Gamma(A)$, for the system of charged particles is given by functionally 
integrating the fields $a_\mu$ and $\xi^\mu_k$ weighted by the lagrangian 
density:
\eq
\label{pL}
\Scl(\xi,a,A) = +\, {1\over 2} \epsilon^{\mu\nu\lambda}
a_\mu \partial_\nu a_\lambda + \Scl_{\rm kin}(\xi) + j^\mu(\xi) (a+A)_\mu,
\eeq
where we have chosen $\theta = - \pi$ and we
need not be concerned about the detailed form of the particle
kinetic term, $\Scl_{\rm kin}(\xi)$, or current, $j^\mu(\xi)$.

On the other hand, the response function of the system related by 
particle-vortex duality to the original system involves integrating
the dual lagrangian
\eqa
\label{vL}
\tilde\Scl(y,a,b,A) &=& -\, {1\over 2} \epsilon^{\mu\nu\lambda}
a_\mu \partial_\nu a_\lambda  -  \epsilon^{\mu\nu\lambda}
b_\mu \partial_\nu (a+A)_\lambda \nonumber\\
&& \qquad + \tilde\Scl_{\rm kin}(y) + 
\tilde{j}^\mu(y) b_\mu,
\eeqa
where we have instead chosen to represent fermionic statistics
by choosing $\theta = + \pi$. In order to relate $\tilde\Gamma(A)$
to $\Gamma(A)$ we explicitly perform the Gaussian integration over $a_\mu$
in eq.~\pref{vL} and shift $b_\mu \to b_\mu + A_\mu$ in the result,
to get
\eqa
\label{vL2}
\tilde\Scl(y,b,A) &=& +\, {1\over 2} \epsilon^{\mu\nu\lambda}
b_\mu \partial_\nu b_\lambda - \frac12 \, \epsilon^{\mu\nu\lambda}
A_\mu \partial_\nu A_\lambda \nonumber\\
&& \qquad  + \tilde\Scl_{\rm kin}(y) + \tilde{j}^\mu(y) 
(b+A)_\mu .
\eeqa

To the extent that $\tilde\Scl_{\rm kin}(y)$ and $\Scl_{\rm kin}(\xi)$
(and $\tilde{j}^\mu(y)$ and $j^\mu(\xi)$) have the same form,
eq.~\pref{vL2} differs from eq.~\pref{pL} only by the term
$-\frac12 \,  \epsilon^{\mu\nu\lambda} A_\mu \partial_\nu A_\lambda$, which
does not depend at all on the integration variables. Performing the 
remaining integrations therefore relates $\tilde\Gamma(A)$ to $\Gamma(A)$ by:
\eq
\label{nonlindual}
\tilde\Gamma(A) = \Gamma(A) - \frac12 \int d^3x \; \epsilon^{\mu\nu\lambda}
A_\mu \partial_\nu A_\lambda ,
\eeq
which is the  main result of this section. 

Once specialized to the linear response regime, eq.~\pref{nonlindual}
implies the relation:
\eq
\label{nlduallr}
\tilde\Pi^{\mu\nu}(p) = \Pi^{\mu\nu}(p)+ i\epsilon^{\mu\lambda\nu} p_\lambda,
\eeq
or: $\tilde\sxy = \sxy + 1$. This agrees exactly with what is predicted 
for fermions by 
a particle-vortex 
transformation followed by a $2\pi$ statistics shift, equation
\pref{dualcondf} followed \pref{2pishiftcond}.

\section{Applications to Conductors}
In this section we state some of the observable predictions which follow from
the action of particle-vortex duality on the physical systems. We specialize
in this section to the predictions for two kinds of conducting systems, 
those with fermionic charge carriers (or their images under repeated duality
transformations) -- corresponding to Quantum Hall systems -- and those with
bosonic charge carriers (or their duality images) -- such as for 
superconducting thin films and/or Josephson junction arrays. 

\subsection{Fermions: The Quantum Hall Effect}
We start with Quantum Hall systems, for which the results
we derive are not new, having been derived from the
the assumption of duality-invariant flows in ref.~\cite{Semicircle}. We
include this case anyway for three reasons. First, this paper strengthens
the theoretical foundation of the assumption of duality-invariant flow,
particularly away from the systems' critical points. Second, the
experimental success of these predictions establishes the existence
of systems which are clean enough for our quasi-particle/vortex effective
theory to apply. Third, we may directly adapt many of the Quantum Hall
results to the bosonic case, for which most of our predictions are new.

\subsubsection{Some Group Theoretical Facts}
In previous sections we have found the action of particle-vortex duality
and $2\pi$ statistics addition to both act on the complex conductivity in
a fractional linear way, with integer coefficients. That is, we have found
the transformations to be a subgroup of the group $PSL(2,Z)$, defined by:
\eq
\label{psl2zdef}
\sigma \to {a \, \sigma + b \over c \, \sigma + d} ,
\eeq
with $a,b,c$ and $d$ integers satisfying $ad-bc=1$. Any element of this
group can be obtained as products of powers of the following two generators:
\eq
\label{sl2zgen}
S(\sigma) := - \; {1 \over \sigma}, \qquad T(\sigma) :=  \sigma + 1.
\eeq

\vtop{
\includegraphics{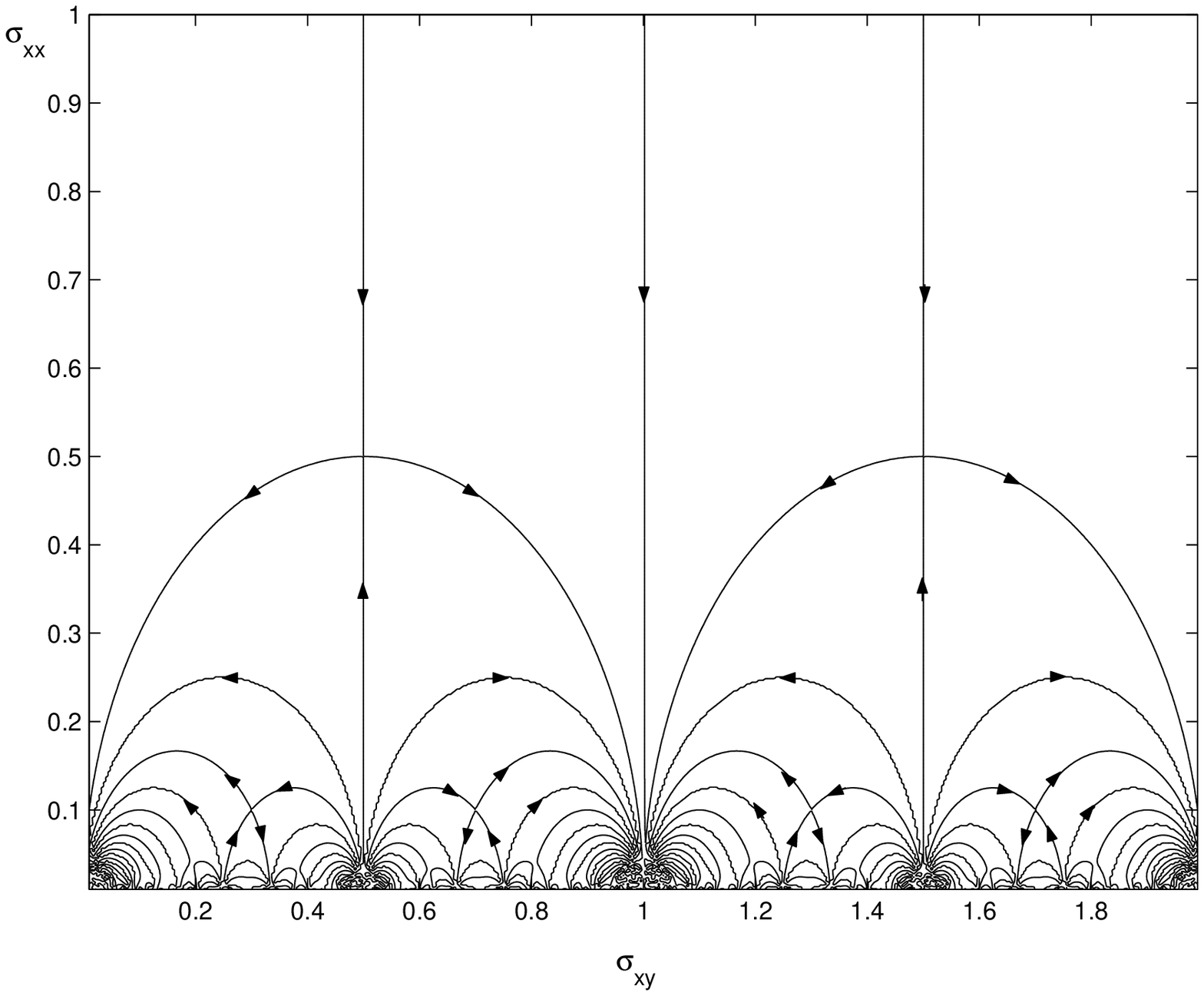}
\vskip 5.5cm
\centerline{Figure 1: Flow lines for $\Gamma_0(2)$-invariant flow.}}
\bigskip
We have found the subgroup of this group which is  
relevant to Quantum Hall systems to be generated by:
\eqa
\label{QHEgen}
\sigma &\to& {\sigma \over 1 - 2\sigma } = ST^2S(\sigma), \nonumber\\
\sigma &\to& {1 - \sigma \over 1 - 2\sigma } = TST^2S(\sigma), 
\eeqa
or, equivalently, by $ST^2S$ and $T$. Any point in the upper complex
$\sigma$-plane can be reached from a fundamental region, which we take
to be the vertical strip between $\sigma=0$ and $\sigma=1$, with the
interior of the disc with radius=$\frac12$ centred at $\sigma = \frac12$ removed. As we
shall shortly see, the boundaries of this fundamental domain are  quite
generally flow lines for $\Gamma_0(2)$-invariant flow, and so may be seen
in Fig.~(1). 

It turns out that the subgroup obtained from these two generators, denoted
$\Gamma_0(2)$, is equivalent to that defined by requiring the coefficient 
$c$ in eq.~\pref{psl2zdef} to be even \cite{Rankin,LutkenRoss,Lutken}. This 
alternative definition shows that when acting on the real axis, 
$\sigma = \sigma^*$, it takes rational numbers to themselves, with
odd-denominator fractions going to odd-denominator fractions, and even-denominator
fractions mapping to even-denominator fractions.\footnote{It is tantalizing
that the same group, $\Gamma_0(2)$, plays a central role in the hierarchical 
structure of $N=2$ supersymmetric Yang-Mills theories \cite{SW}.}

\subsubsection{Predictions}
The following consequences follow from the condition that a flow in the
$\sigma$-plane commutes with the group $\Gamma_0(2)$:
\begin{enumerate}
\item
Universal critical points \cite{UnivcondFisheretal}
are predicted for the flow at any
point, $\sigma_{crit}$, which is mapped to itself under any member of 
$\Gamma_0(2)$, $\gamma(\sigma_{crit}) = \sigma_{crit}$, for which the denominator
$c\, \sigma + d$ is neither zero or 
infinite \cite{LutkenRoss,Lutken,LutkenBurgess}. The complete set of
such points occur at $\sigma_{crit} = \frac12 (1+i)$ and its images under $\Gamma_0(2)$.
This prediction is borne out experimentally \cite{critsig},
since there is a one-to-one correspondence between the experimental 
critical points of the flow and the fixed points of the group. No experimental 
evidence exists for critical points not corresponding to fixed points of the 
group, although these could exist in principle.
\item
Although the symmetry does not predict the critical exponents at
the fixed points \cite{B+L}, these exponents must be the same for
all fixed points which are related by $\Gamma_0(2)$
 \cite{LutkenRoss,Lutken,LutkenBurgess}. This remarkable equivalence
of critical exponents at different fixed points is 
known as super-universality, and was argued for on more model-dependent
grounds in ref.~\cite{SUnivmicroargs}. 
Historically, the experimental success of this
`prediction' \cite{SUnivWTPP,SUnivEngel} 
stimulated the search for an underlying symmetry group.
\item
Exact flow lines in the $\sigma$ plane can be derived from $\Gamma_0(2)$
invariance plus invariance under particle-hole symmetry: $\sigma \to 
1 - \sigma^*$. The existence and shape of these flow lines depends only
on these symmetries and not on any other details of the flow's $\beta$ 
function \cite{Semicircle}. Figure (1) shows some of the flow lines 
predicted in this way, all of which are semi-circles or vertical lines
in the $\sigma$-plane. The arrows indicate the direction of flow
to the infrared (whose direction does not follow purely on symmetry
grounds). This very general derivation of the 
`semi-circle law' -- which had been earlier predicted on more
model-dependent grounds \cite{SCRuzin} -- is spectacularly exhibited
by experimental systems \cite{SCHilkeetal}. 
\item
Flow in the infrared is towards the real axis, terminating on the
real axis at attractive fixed points at odd-denominator fractions. 
Even-denominator fractions similarly form repulsive fixed points
of the flow. One finds in this way a robust explanation of the existence
of odd denominator Hall plateaux. 
\item
Since all allowed transitions between Hall plateaux correspond to
semi-circles which may be obtained by the action of $\Gamma_0(2)$
from the basic semi-circle connecting $\sigma = 0$ to $\sigma = 1$,
one finds a selection rule which expresses which plateaux may be
obtained from which by varying external parameters like magnetic
field and temperature \cite{BD}. The selection rule obtained in this 
way states that a fraction $p_2/q_2$ can be reached by a quantum Hall 
transition from a fraction $p_1/q_1$, with $q_1$ and $q_2$ both odd integers, 
only if $|p_1 q_2 - p_2 q_1 | =1$. This agrees precisely 
with all the observed Hall sequences.
\item
There is an element of $\Gamma_0(2)$ which maps each of the flow
lines to itself, with its ends reversed. For instance, the element
$\gamma(\sigma) = (\sigma-1)/(2\sigma - 1)$, which we here identify
as the expression of particle-vortex duality for fermions, does so for 
the semi-circle connecting $\sigma = 0$ to $\sigma = 1$. When specialized to 
transitions from Laughlin plateaux to the Hall insulator this 
symmetry element is precisely the experimentally observed \cite{dualSTSCSS}
duality $\rho_{xx} \to 1/\rho_{xx}$ with $\rho_{xy}$ fixed \cite{Semicircle}.
\end{enumerate}

We regard the natural interpretation of the great experimental success 
of these predictions to be that the electromagnetic response of these
systems is dominated by quasi-particles and vortices. Furthermore
the systems are sufficiently clean to justify the neglect of those 
particle and vortex interactions which destroy the underlying particle-vortex
duality, as described here. Finally, the relevant quasi-particles are 
in the fermionic equivalence class, in the sense that they are either
fermions, or are obtainable from fermions by the action of $\Gamma_0(2)$.

There is also some evidence for a few Hall systems for which the
critical conductivity is not at the universal values \cite{SCHilkeetal}, 
and for direct plateau-insulator transitions which do not correspond to 
semi-circles as predicted here \cite{wrongplateaux}, and an understanding
of why particle-vortex duality fails here would be very instructive. 
Since these typically involve samples with the most disorder, one possibility
is that Landau-level mixing is not negligible in these 
systems \cite{Hilkepriv,Shimshoni}. 
We believe this to an instance where interactions with the 
disorder ruin particle-vortex duality,
and so destroy the underlying symmetry of the flow.

Another potential difficulty often raised in connection with this picture is 
the observed failure of scaling at very low temperatures in some samples \cite{noscale}
as one passes through the critical regime. We put these experiments aside,
since although these are potentially very telling experiments,
since scaling is an inevitable consequence of a vanishing $\beta$ function, it is 
not yet clear what their proper interpretation is, and indeed there are other
experiments which appear to support scaling, \cite{Coleridge}.\footnote{See, 
however, ref.~\cite{Crossover} for an
alternative interpretation of the behaviour near the critical points.}

There is nevertheless content in the above symmetry arguments, since these
imply an entire suite of predictions which
must all hold together if particle-vortex duality is valid. So we
predict that the above consequences of $\Gamma_0(2)$-invariant flow should 
come as a package, with the failure of some implying the failure
of the others.

\subsection{Bosons: Superconducting Films}
A fascinating consequence of the generality of the particle-vortex duality
arguments we present here is that they predict different, but equally
striking, phenomena for the electromagnetic response of 
other clean two-dimensional systems. In this section we describe
these predictions for systems whose charge-carrying quasi-particles have
bose statistics (or the image of bose statistics under a group which
we here specify). These predictions should have practical applications to
superconducting thin films and Josephson Junction arrays, and some
of them have been anticipated \cite{Fisher,FisherLee,LeeKane} 
for the metal-insulating and superconductor-insulator transitions of 
these systems. 

\subsubsection{More Group Theoretical Facts}
For bosonic charge carriers (and those related to these by duality)
the action of particle-vortex duality and $2\pi$ statistics 
addition is generated by the following two $PSL(2,Z)$ elements:
\eqa
\label{boxgen}
\sigma &\to& {\sigma \over 1 - 2\sigma } = ST^2S(\sigma), \nonumber\\
\sigma &\to& - \; {1\over \sigma } = S(\sigma) ,
\eeqa
or, equivalently, $S$ and $T^2$. 

This group is called $\Gamma_\theta(2)$, and is equivalent to the 
condition that $a,d$ are both odd and $b,c$ are both even, or vice versa,
in the fractional linear transformation, $(a\, \sigma + b)/(c\, \sigma +d)$.
Since bosons may be obtained from fermions by shifting their statistics
by $\Delta \theta = \pi$, this group may be obtained from the Quantum 
Hall group, $\Gamma_0(2)$, by conjugating by $STS(\sigma) = \sigma/(1-\sigma)$. 
Concretely: $g \in \Gamma_\theta(2)$ implies $g = STS \, h \, (STS)^{-1}$, 
for some $h \in \Gamma_0(2)$. (The proof of this statement is easiest to see if the
identity $(ST)^3 = 1$ is used.) The simplest way to extract the predictions 
of the group $\Gamma_\theta(2)$ is therefore to derive them from those 
of $\Gamma_0(2)$ by conjugating with $STS$. (Or, since $S$ is in 
$\Gamma_\theta(2)$ anyway, we can equally well get $\Gamma_\theta(2)$ by 
conjugating $\Gamma_0(2)$ with $TS(\sigma)=1-{1\over \sigma}$ rather than $STS$.) 
In particular, it is convenient
to choose the fundamental region to be the vertical strip between 
$\sigma=0$ and $\sigma=-1$, with the interior of the disc with radius=$\frac12$ centred at 
$\sigma = -\, \frac12$ removed, the boundaries of which again appear as 
particular flow lines in figure (2). 

\vtop{
\includegraphics{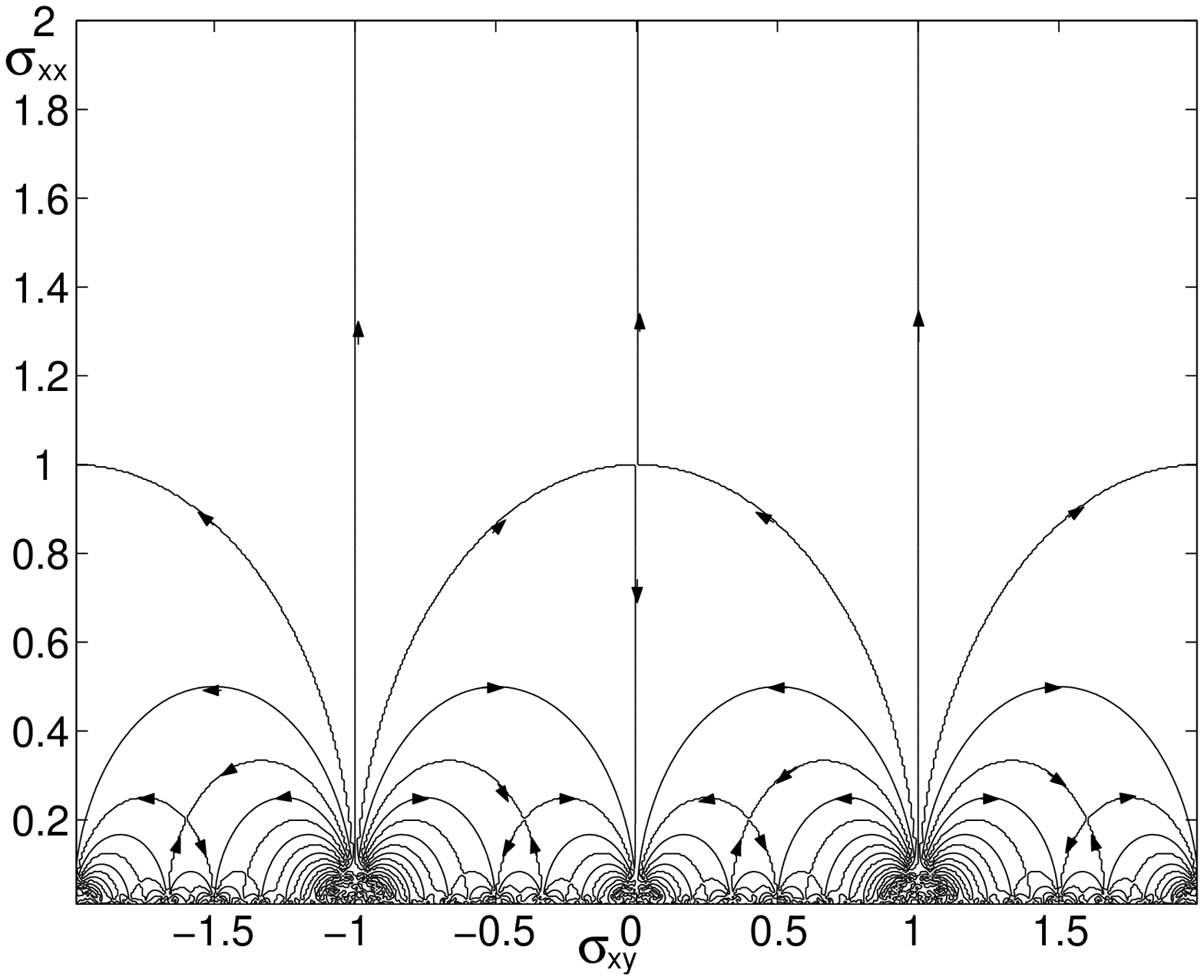}
\vskip 5.5cm
{Figure 2: Flow lines for $\Gamma_\theta(2)$-invariant flow. Although
this looks very much like Figure 1, careful examination shows different positions
for the fixed points, and for the directions of flow along the lines.}}

\subsubsection{Predictions}
In this way we obtain the following consequences of the commuting of 
$\Gamma_\theta(2)$ with the flow in the $\sigma$-plane, which are the
direct analogs of those described above for Quantum Hall systems.
\begin{enumerate}
\item
Universal critical points are predicted for the flow at the fixed points
of $\Gamma_\theta(2)$, which lie at $\sigma_{crit} = i$ and its images,
$(bd + ac + i)/(c^2 + d^2)$, under $\Gamma_\theta(2)$. 
The Ohmic conductivity is always $1/(\hbox{odd integer})$ at these critical points
while the Hall conductivity is
$\hbox{even}/\hbox{odd}$.
(These statements are for bosonic charge carriers with the
same electric charge as an electron --- 
these become $\sigma_{crit} = i q^2$ and its images if the bosonic charge
carriers have charge $q$. In particular, the case $q=\pm 2$ applies if
the bosons are Cooper pairs, such as considered in ref.~\cite{Fisher}.) 
\item
The critical exponents at all fixed points related by $\Gamma_\theta(2)$
must all be the same. In this way the results of the scaling theory
of ref.~\cite{Fisher} for the fixed point at $\sigma_{crit}=i$ may be 
extended to all of the other fixed points which are predicted to
exist in the presence of magnetic fields.
\item
Exact flow lines in the $\sigma$ plane are immediate consequences of 
$\Gamma_\theta(2)$ invariance and particle-hole symmetry, independent
of the dynamical details of the flow's $\beta$ function. These
are the images of the flow lines of Figure (1) under the conjugation
by $STS$. The results are again semi-circles or vertical lines
in the $\sigma$ plane, implying a new semi-circle law for these
bosonic systems. (The semi-circle intersecting the critical point
at $\sigma = i$ was anticipated in ref.~\cite{Fisher}.)

Figure (2) shows the flow lines which are predicted by the symmetry,
with the flow directions given which follow from those of Figure (1)
(which successfully describe Quantum Hall systems). Notice that
the resulting flow on the imaginary axis agrees with the interpretation
of $\sigma = i$ as a metal-insulator transition, with $\sxx$ increasing
or decreasing into the infrared on opposite sides of the transition.
\item
More generally, for nonzero magnetic fields the flow in the 
infrared is towards the real axis, terminating on the attractive 
fixed points which are fractions, $\sigma = p/q$. The attractive
fixed points of the flow therefore exhibit the fractional QHE,
but with fractions for which $pq$ is even (as opposed to having
$q$ odd, as was the case for fermions). Fractions with odd $pq$ 
are repulsive fixed points. In particular odd integers are repulsive
while even integers are attractive.
\item
There is a selection rule expressing which fractions may be
obtained from which by varying external parameters like magnetic
field and temperature. The selection rule obtained in this way states that 
fractions $p_2/q_2$ can be obtained from $p_1/q_1$ only if: ($i$) $p_1$ is odd 
and $q_1$ is even while $p_2$ is even and $q_2$ is odd (or vice versa with the
subscripts 1 and 2 interchanged) and ($ii$) $|p_1 q_2 - p_2 q_1|=1$.
\item
There is an element of $\Gamma_\theta(2)$ which maps flow
lines to themselves, with their ends reversed. For flow along
the imaginary axis the element is simply $S(\sigma) = -1/\sigma$,
or $\sigma_{xx} \to 1/\sigma_{xx}$, 
which is again the expression of particle-vortex duality for bosons. 
\end{enumerate}

We are led to predict the above startling properties for bosonic Quantum Hall systems,
such as someday might be obtained from superconductor-insulator transitions in thin films, or from 
Josephson-junction arrays \cite{JJtheory0,JJtheory1,JJtheory2,JJtheory3,JJtheory4}. 
Unfortunately, these predictions cannot yet be tested with ordinary superconductors,
because charge carrier densities and mobilities in these systems
do not put them into the quantum Hall regime \cite{HiTcmob}.
Their verification would be the smoking gun for particle-vortex 
duality, although its testing must wait until 
these systems can be reliably manufactured in the quantum regime. In this way
particle-vortex duality might ultimately provide a connection between Quantum Hall
systems and $\sxx \to 1/\sxx$ as seen in 
metal-insulator transitions \cite{MIduality1,MIduality2}. 

As in the case of Quantum Hall systems, these predictions require very few
assumptions beyond the necessity to be in the quantum regime. 
First, the electromagnetic response must be dominated by 
quasi-particles and vortices, with the quasi-particles being bosons or
related to these by $\Gamma_\theta(2)$ transformations. Second, the
systems must be sufficiently clean to ensure the absence of interactions
which distinguish the quasi-particles from vortices, and so ruin their
similarity. Subject to these conditions all of the above consequences
follow as a package from the quasi-particle-vortex duality symmetry
of the flow in the conductivity plane. 

\section{Conclusions and Outlook}

We have argued that a broad class of dual relationships arise in two dimensional
systems for which the EM response is governed by particles and vortices whose 
properties are similar (perhaps because they are weakly interacting). For systems
having fermions as the particles (or those related to fermions by the duality)
the particle-vortex duality implies the duality group is a
level-two subgroup of $PSL(2,Z)$ called $\Gamma_0(2)$. We argue that this duality
has been observed, since this group has been previously identified as explaining many 
of the observed properties of the EM response in Quantum Hall systems. 

The generality of our arguments lead us to propose a similar, but distinct, set of properties for
clean two-dimensional systems whose EM response is controlled by bosons as the particles (or those
related to bosons by the duality). The group implied in this case is another
level-two subgroup of $PSL(2,Z)$ called $\Gamma_\theta(2)$. The observation of these
specific predictions in these systems would be the definitive test of our ideas.

One might ask whether other duality symmetries apart from those described above might arise in 
other systems. More concretely, there are precisely five level-two subgroups of 
$SL(2,Z)$ \cite{Rankin}, so one might wonder if other choices for quasi-particle 
statistics might generate these other three groups not yet used \cite{andynpb}. 
In fact, two of these groups can be ruled out as symmetries acting on the complex
conductivity, because they do not contain the generator corresponding to the addition of
$2\pi$ statistics flux. However, because these groups are related to $\Gamma_0(2)$  and
$\Gamma_\theta(2)$ by conjugating by $S$, they can be thought of as the action of these
latter groups in the complex resistivity, rather than conductivity, plane. The third group,
$\Gamma(2)$, is contained in the other four, and has been proposed elsewhere to play a role for 
Quantum Hall systems  \cite{GMW}, in particular 
when the splitting between electron spins is much smaller than the
gap between successive Landau levels \cite{BDspin}. 

\begin{center}
{\bf ACKNOWLEDGMENTS}
\end{center}

We thank M. Hilke, C.A. L\"utken, M. Paranjape, F. Quevedo and S. Sondhi 
for helpful discussions. C.B. is grateful to the University of Barcelona, D.A.M.T.P 
at Cambridge University, and the Aspen Centre for Physics, and B.D. to the 
Groupe de Physique Th\'eorique, IPN, Orsay, and Dept. de Fisica, CINVESTAV, 
Mexico City, Mexico for their kind hospitality while part of this work was carried out.
Our research has been assisted by financial support from N.S.E.R.C. (Canada), F.C.A.R. 
(Qu\'ebec), the Ambrose Monell Foundation, CNRS (France) and Enterprise 
Ireland, Basic Research Grant no. SC/1998/739.

\bigskip
\begin{center}
{\bf APPENDIX A\\
The Statistical Gauge Field}
\end{center}

In this appendix we review the concept of the statistical gauge
field \cite{ASW}.  This serves not only to remind the reader of the construction, 
but also to set up the notation. This appendix is based on the review
article by S-C.~Zhang, \cite{Zhang}.

\topic{First-Quantized Formulation}

Consider a configuration of $N$ charged particles, each with the same 
charge $e$ and mass $m$ moving in a two dimensional
plane, with positions  ${\bf x}_1,\dots,{\bf x}_N$ and
separations ${\bf r}_{ij}={\bf x}_i-{\bf x}_j=-{\bf r}_{ji}$.
Allowing for particle-particle interactions, $e^2V({\bf x}_i-{\bf x}_j)$,
and the possibility of random static impurities giving, rise to an electric potential
$U({\bf r})$, the Hamiltonian for the system can be written as:
\begin{eqnarray}
H={1\over 2m}\sum_i\left(-i\hbar\nabla_\alpha^{(i)}-eA_\alpha({\bf x}_i\right)^2
&+e^2\sum_{i<j}V({\bf x}_i-{\bf x}_j)\nonumber\\
+e\sum_iU({\bf x}_i),&
\end{eqnarray}
where $\alpha=1,2$ and we use units in which the speed of light $c=1$.
One can include a neutralising background field if desired, 
without changing any
of the subsequent analysis significantly.

Following MacDonald \cite{MacDonald}, a gauge transformation to a new Hamiltonian is defined
as follows.
Let $\varphi_{ij}$ be the angle between the vector ${\bf r}_{ij}$ and an
arbitrary fixed direction, 
{\it e.g.}~the $x$-axis, so that $\varphi_{ij}\rightarrow\varphi_{ij}+\pi$
if the two particles $i$ and $j$ are interchanged. A gauge transformation from the
multi-particle Schr\"odinger wave-function, $\psi({\bf x}_1,\dots,{\bf x}_n)$
to a new wave-function $\tilde\psi({\bf x}_1,\dots,{\bf x}_n)$
is defined as follows,
\begin{equation}
\label{GT}
\tilde\psi({\bf x}_1,\dots,{\bf x}_N)=
\hbox{e}^{i{\theta\over\pi}\sum_{i<j}\varphi_{ij}}
\psi({\bf x}_1,\dots,{\bf x}_N)
\end{equation} 
for a constant $\theta$, as yet arbitrary.
Under interchange of any two particles, $i$ and $j$, the phase factor changes
by $\hbox{e}^{i\theta}$. Thus, if $\theta=2k\pi$ for some integer $k$, the phase factor
is unity and the new wave function has exactly the same statistics as the old one ---
if it was symmetric under interchange of two particles before it remains so, if
it was anti-symmetric under interchange of two particles before it remains so. 
If $\theta$ were an odd multiple of $\pi$ a bosonic wave function would be transformed
into a fermionic one and vice versa.

The gauge transformation \pref{GT} can be incorporated into the vector potential $A_\mu$, 
$\mu=0,1,2$, for the electro-magnetic field by defining a new field (the statistical
gauge field)
\begin{equation}
\label{eqthree}
a_\alpha({\bf x}_i):=
{\hbar\over e}{\theta\over\pi}\sum_{\{j;j\ne i\}}\nabla_\alpha^{(i)}\varphi_{ij}
\end{equation} 
where the gradient operator acts on the position of the $i$-th particle.
The new gauge transformed Hamiltonian is then
\begin{eqnarray}
\tilde H={1\over 2m}\sum_i&\left(-i\hbar\nabla_\alpha^{(i)}-eA_\alpha({\bf x}_i)
-ea_\alpha({\bf x}_i)\right)^2\nonumber\\
&{+e^2\sum_{i<j}V({\bf x}_i-{\bf x}_j)+e\sum_iU({\bf x}_i)}.
\end{eqnarray}
It is stressed that the physics of this new Hamiltonian is identical to the old
one, provided ${\theta\over 2\pi}$ is an integer.

\topic{Second-Quantized Formulation}

We now reformulate the above first quantised treatment in second quantised
form,
ignoring spin.
Accordingly define,
\begin{eqnarray}
{\tilde{\bf H}=}&\nonumber\\
&\kern -1cm\int d^2x\Psi^\dagger({\bf x})
\left[{1\over 2m} \bigl(-i\hbar{\hbox{\Vt{$\nabla$}}}
\kern -3pt -e{\bf A}({\bf x})-e{\bf a}({\bf x})\bigr)^2
\kern -2pt +eU({\bf x})\right]\Psi({\bf x})\nonumber\\ 
&\kern -2cm +{1\over 2}\int d^2xd^2x^\prime
\delta\rho({\bf x})V({\bf x}-{\bf x}^\prime)\delta\rho({\bf x}^\prime),
\end{eqnarray}
where $\rho({\bf x})=e\Psi^\dagger({\bf x})\Psi({\bf x})$ is the charge density
and \hbox{$\delta\rho=\rho-<\rho>$}.

Equation \pref{eqthree} can be expressed as
\begin{equation}
a_\alpha({\bf x}_i)=
-\left({\hbar\theta\over e\pi}\right)
\sum_{j\ne i}\epsilon_{\alpha\beta}{({\bf r}_i-{\bf r}_j)^\beta\over\vert{\bf r}_i-{\bf r}_j\vert^2}
\end{equation}
which in continuum form is
\begin{equation}
a_\alpha({\bf x})=
-\left({\hbar\theta\over e^2\pi}\right)\int d^2x^\prime 
{\epsilon_{\alpha\beta}\bigl(x^\beta-x^{\prime\beta}\bigr)\over
\vert {\bf x}-{\bf x}^\prime\vert^2}
\rho({\bf x}^\prime).
\end{equation}
>From this follows
\begin{equation}
\label{spacecurla}
\epsilon^{\beta\alpha}\nabla_\beta a_\alpha({\bf x})=
{2\hbar\theta\over e^2}\rho({\bf x}),
\end{equation}
since $\nabla^2\ln(\vert {\bf x}-{\bf x}^\prime\vert)
=2\pi\delta^{(2)}(\vert {\bf x}-{\bf x}^\prime\vert)$.
The curl of ${\bf a}$ is non-zero here, despite it's original
definition as a gradient, because of Aharanov-Bohm type singularities
in \pref{eqthree}. These manifest themselves as $\delta$-function singularities
in the curl of ${\bf a}$ in the first quantised theory but are
spread into a smooth distribution on the right hand side of \pref{spacecurla},
in the second quantised theory.
Equation \pref{spacecurla}  is a dynamical equation of motion for the statistical gauge field $a_\mu$,
but it is not yet co-variant, as it lacks $a_0$ terms. To include these
allow $a_\alpha$ and $\rho$ to depend on time, then
differentiating \pref{spacecurla} and expressing the result co-variantly gives
\begin{equation}
\epsilon^{\beta\alpha}\partial_\beta (\partial_0a_\alpha-\partial_\alpha a_0)
={2\hbar\theta\over e^2}\dot\rho=- {2\hbar\theta\over e^2}\partial_\alpha j^\alpha,
\end{equation}
where $j^\alpha$ is a current.
This integrates to 
\begin{equation}
\epsilon^{\alpha\beta}(\partial_0 a_\beta -\partial_\beta a_0)=
-{2\hbar\theta\over e^2}j^\alpha
\end{equation}
This equation can be obtained, together with equation \pref{spacecurla}, by treating $a_\mu$
as a dynamical field in the action
\begin{equation}
S=-\left({e^2\over 4\hbar\theta}\right)
\int dtd^2x \epsilon^{\mu\nu\lambda}
a_\mu\partial_\nu a_\lambda +\int dtd^2x a_\mu j^\mu,
\end{equation}
which is the Chern-Simons action for the statistical gauge field coupled to a source.

The problem can now be re-written in path integral form.
Define the original Lagrangian density for the matter fields
\begin{eqnarray}
{\mathcal L}_\Psi&=&
\Psi^\dagger
\bigl(i\hbar\partial_t-eA_0\bigr)\nonumber\\
&&-\Psi^\dagger
\left[{1\over 2m} \bigl(-i\hbar{\hbox{\Vt{$\nabla$}}}
-e{\bf A}\bigr)^2
-eU\right]\Psi\nonumber\\ 
&&\qquad\qquad-{1\over 2}\int d^2x^\prime
\delta\rho({\bf x})V({\bf x}-{\bf x}^\prime)\delta\rho({\bf x}^\prime).
\end{eqnarray}
After introducing the statistical gauge field the new
Lagrangian is
\begin{eqnarray}
\label{LAa}
\tilde{\mathcal L}_\Psi&=&\Psi^\dagger
\bigl(i\hbar\partial_t-e(A_0+a_0)\bigr)\Psi\nonumber\\
&&-\Psi^\dagger
\left[{1\over 2m} \bigl(-i\hbar{\hbox{\Vt{$\nabla$}}}
-e{\bf A}-e{\bf a}\bigr)^2
-eU\right]\Psi\nonumber\\ 
&&\qquad -{1\over 2}\int d^2x^\prime
\delta\rho({\bf x})V({\bf x}-{\bf x}^\prime)\delta\rho({\bf x}^\prime),
\end{eqnarray}
together with the Chern-Simons Lagrangian
\begin{equation}
{\mathcal L}_a=-\left({e^2\over 4\hbar\theta}\right)
\epsilon^{\mu\nu\lambda}
a_\mu\partial_\nu a_\lambda.
\end{equation}
Then the effective action for the gauge field $A_\mu$
is obtained, in the original formalism, from the path integral
\begin{equation}
\label{ZA}
\hbox{e}^{i\Gamma_{eff}[A]}=Z[A]=\int{\mathcal D}\Psi^\dagger{\mathcal D}\Psi
\hbox{e}^{iS_\Psi[A_\mu,\Psi^\dagger,\Psi
]},
\end{equation}
where $S_\Psi=\int dtd^2x{\mathcal L}_\Psi$.
On the other hand, after introducing the statistical gauge field,
the partition function is
\begin{eqnarray}
\label{ZAa}
\tilde Z[A]&=&\int{\mathcal D}a_\mu{\mathcal D}\Psi^\dagger{\mathcal D}\Psi
\hbox{e}^{iS_a[a_\mu]+iS_\Psi[A_\mu+a_\mu,\Psi^\dagger,\Psi
]}\nonumber\\
&=&\int{\mathcal D}a_\mu \hbox{e}^{iS_a[a_\mu]}Z[A+a],
\end{eqnarray}
where $S_a=\int dtd^2x{\mathcal L}_a$.

So far all manipulations have been exact. For $\theta=2k\pi$,
with integral $k$, the theory described by \pref{ZAa} is
{\it identical} to that with no statistical gauge field at all,
i.e. $\tilde Z[A]={\mathcal N}Z[A]$, where ${\mathcal N}$ is an irrelevant
constant that will be ignored in the sequel.

\bigskip
\begin{center}
{\bf APPENDIX B\\
The Relativistic Trick}
\end{center}

We now follow Kivelson, Lee and Zhang \cite{KLZ} and integrate out the matter
fields. This cannot be done exactly, of course, but on general grounds
one expects the effective action to be invariant
under local gauge transformations. 
Firstly ignore the statistical gauge field and 
consider the partition function (\ref{ZA}).
Na{\rm\"\i}ve power counting arguments
then imply that the most relevant terms, at least in the infra-red limit,
are
\begin{eqnarray}
\label{origaction}
\Gamma_{eff}[A]&=&\nonumber\\
&&\kern -1cm \int d^3xd^3x^\prime\left\{
\left({1\over 2}\right)
F_{i0}(x)\pi_1(x-x^\prime)F_{i0}(x^\prime)\right.\nonumber\\
&&-\left({1\over 2}\right)F_{12}(x)\pi_2(x-x^\prime)F_{12}(x^\prime)\nonumber\\
&&\left.-{1\over 2}\epsilon^{\mu\nu\lambda}A_\mu(x)\pi_3(x-x^\prime)\partial_\nu 
A_\lambda(x^\prime)
\right\}
\end{eqnarray}
where $F_{i0}=E_i$, $i=1,2$, is the electric field and $F_{12}=B$ the magnetic
field (the measure here $d^3x$ is a shorthand for $dtd^2x$ and
Greek indices, $\mu,\nu,\lambda$ take three values, $0,1$ and $2$).
Note the presence of the non-local form factors $\pi_1,\pi_2$ and $\pi_3$,
where the argument $x-x^\prime$ includes time as well as space.

There is an implicit assumption here, that the terms that are most
relevant by na{\rm\"\i}ve power counting are also the most relevant in the 
full theory, at least in the infra-red limit. This is a very strong assumption, 
as na{\rm\"\i}ve
power counting gives the most relevant operators of the free field 
theory and we have here a strongly interacting theory. One of 
the underlying
assumptions of Kivelson, Lee and Zhang's argument is therefore that
any anomalous dimensions in the strongly interacting theory do
not change the conclusions of na{\rm\"\i}ve power counting.
In fact the form (\ref{origaction}) is rather
more general than that --- it will be argued below that this
is the most general possible form in a momentum expansion, independently
of any power counting arguments, provided one allows the polarizations to
depend on the magnitudes of the external fields, ${\bf E}^2$ and ${\bf B}^2$.

The mathematical manipulations involving equation (\ref{origaction})
can be simplified by writing it in
relativistic form using the following trick.
We introduce a \lq\lq metric" on 3-dimensional space-time defined by
\begin{equation}
\label{metric}
g_{\mu\nu}(x-x^\prime)= \left(\matrix{-1&0&0\cr
0&{\pi_1\over\pi_2}&0\cr
0&0&{\pi_1\over\pi_2}\cr}\right)
\end{equation}
and write 
\begin{eqnarray}
\label{origactionm}
\Gamma_{eff}[A]&=&\nonumber\\
&&\kern -1.5cm \int d^3xd^3x^\prime\left\{
-{1\over 4}\pi_1(x-x^\prime)\sqrt{-det(g)}
g^{\mu\rho}g^{\nu\tau}F_{\mu\nu}(x)F_{\rho\tau}(x^\prime)\right.\nonumber\\
&&\left.-{1\over 2}
\epsilon^{\mu\nu\lambda}A_\mu(x)\pi_3(x-x^\prime)\partial_\nu A_\lambda(x^\prime)\right\},
\end{eqnarray}
where all metric components are functions of $x-x^\prime$. Note that the metric does
not appear in the Chern-Simons term at all --- as is well known it is
independent of the metric.

The calculation can be simplified by setting $\pi_1=\pi_2$
and working in the relativistic formalism with the 3-dimensional
Minkowski metric, $\eta_{\mu\nu}=diag(-1,1,1)$, and relativistic action 
\begin{eqnarray}
\label{origactionr}
\Gamma_{eff}[A]&=&\nonumber\\
&&\kern -1cm \int d^3xd^3x^\prime\left\{
-\left({1\over 4}\right)\pi_1(x-x^\prime)
F^{\mu\nu}(x)F_{\mu\nu}(x^\prime)\right.\nonumber\\
&&\left.-{1\over 2}\epsilon^{\mu\nu\lambda}A_\mu(x)\pi_3(x-x^\prime)
\partial_\nu A_\lambda(x^\prime)
\right\},
\end{eqnarray}
where $F^{\mu\nu}=\eta^{\mu\rho}\eta^{\nu\tau}F_{\rho\tau}$.
Provided everything is kept co-variant the non-relativistic
expressions can be recovered by re-instating the metric (\ref{metric}) at
the end.  From now on we shall use the simpler relativistic notation
of equation (\ref{origactionr}).

In relativistic notation we can argue that (\ref{origactionr}) actually 
encodes higher
order terms too. In strong fields one would expect terms  such as
${F^\mu}_\nu{F^\nu}_\rho{F^\rho}_\lambda{F^\lambda}_\mu$, and other Lorentz invariant powers,
to be present in the effective action
even in the low momentum regime --- though terms like
$\partial^2F^{\mu\nu}F_{\mu\nu}$ are definitely ignored.
In three dimensions  terms like ${F^\mu}_\nu{F^\nu}_\rho{F^\rho}_\lambda{F^\lambda}_\mu$ 
can be absorbed into
(\ref{origactionr}) by allowing the polarizations to depend on the Lorentz
scalar $F_{\mu\nu}F^{\mu\nu}$. This is because we can always exchange the
anti-symmetric tensor $F_{\mu\nu}$ for the vector 
${\tilde F}^\mu=\epsilon^{\mu\nu\rho}F_{\nu\rho}$ and the only way to make a Lorentz
scalar from products of ${\tilde F}^{\mu_1}\cdots {\tilde F}^{\mu_n}$ is to contract the indices in
pairs (and $n$ must be even), so all such terms can simply be incorporated
into the statement that $\pi_1$ depends analytically on the Lorentz
scalar $F^2=F^{\mu\nu}F_{\mu\nu}$ 
and then Taylor expanding $\pi_1$ in 
powers of $F^2$.
\footnote{In general this would necessitate the 
introduction of multi-point
interactions with $n$ points, $x_1,\ldots,x_n$, but in the long wavelength
limit two point interactions suffice to extract conductivities.} 
An exactly similar statement 
applies to $\pi_3$. We shall continue with the form 
(\ref{origactionr}), bearing in mind that when 
the external fields are strong the polarizations may depend
on them explicitly.

It will be more convenient to work in momentum space where 
(\ref{origactionr}) reads
\begin{eqnarray}
\label{origactionp}
\Gamma_{eff}[A]&=&\int d^3\bar p\left\{
-\left({1\over 4}\right)\pi_1(p)
F^{\mu\nu}(-p)F_{\mu\nu}(p)\right.\nonumber\\
&&\qquad\qquad\left.-{i\over 2\hbar}\epsilon^{\mu\nu\lambda}A_\mu(-p)\pi_3(p) p_\nu  
A_\lambda(p)
\right\},
\end{eqnarray}
with $d^3\bar p={d^3p\over \hbar^3}$.
  
A more compact way of writing
(\ref{origactionp}) is
\begin{eqnarray}
\label{origactionP}
\Gamma_{eff}[A]&=&-\int d^3\bar p\left\{
{1\over 2}\Pi_1(p)
A_\mu(-p)\Lambda^{\mu\nu}A_\nu(p)\right.\nonumber\\
&&\qquad\qquad\left.+{1\over 2}\Pi_3(p)J^{\mu\nu}A_\mu(-p)
A_\nu(p)
\right\},
\end{eqnarray}
where ${\Lambda^\mu}_\nu={\delta^\mu}_\nu -{p^\mu p_\nu\over{p^2}}$, 
$J^{\mu\nu}=i\epsilon^{\mu\lambda\nu}p_\lambda/\sqrt{p^2}$,
$\Pi_1=\pi_1 p^2/\hbar^2$ and $\Pi_3=\pi_3\sqrt{p^2}/\hbar$.
This form is useful, since 
${\Lambda^\mu}_\alpha{\Lambda^\alpha}_\nu={\Lambda^\mu}_\nu$, 
${\Lambda^\mu}_\alpha J^{\alpha\nu}=J^{\mu\alpha}{\Lambda^\alpha}_\nu=J^{\mu\nu}$
and $J^{\mu\alpha}J_{\alpha\nu}=-{\Lambda^\mu}_\nu$, which
makes Gaussian integrals particularly simple.
Note that $J$ is a Hermitian matrix for space-like momenta, but anti-Hermitian
for time-like momenta --- however $\sqrt{p^2}J$ is always Hermitian, which
is all that is necessary to
ensure that the momentum space expression is always Hermitian.

It is stressed that \pref{origactionP} is the most general form of the
effective action in the long-wavelength limit, even for strong fields,
provided $\Pi_1$ and $\Pi_3$ are allowed to depend on $|{\bf E}|^2$ and
$|{\bf B}|^2$.

\begin{center}
{\bf APPENDIX C\\
The Duality Prescription}
\end{center}

The purpose of this appendix is to derive the duality transformation from
eq.~\pref{LELWLagr} to eq.~\pref{dualLagr}, which we do following the general
duality prescription \cite{buscher}. Our starting point is an extended
Lagrangian which is obtained by coupling $\phi$ to a new gauge field, $\Sca_\mu$, 
which is constrained to be pure gauge:
\eqa
\label{ExtLagr}
\Scl_\ext &=& - \, {\pi \over 2 \theta} \; \epsilon^{\mu\lambda\nu} \,
a_\mu \partial_\lambda a_\nu + \Scl_p(\xi,a+A) \nonumber\\
&& - \frac{\kappa}{2} \, [\partial_\mu \phi
- q_\phi (a_\mu + A_\mu + \Sca_\mu)]^2 \nonumber\\
&& + \epsilon^{\mu\nu\lambda} b_\mu \partial_\nu \Sca_\lambda
+ \cdots ,
\eeqa
where the field $b_\mu$ is a Lagrange-multiplier field which is introduced
to enforce the vanishing of the field strength, $\partial_\mu \Sca_\nu 
-\partial_\nu \Sca_\mu$. 

That this extended action is precisely equivalent to the original action
may be seen by first performing the integration over $b_\mu$, which produces
a functional delta function which enforces the constraint 
$\epsilon^{\mu\nu\lambda} \partial_\nu \Sca_\lambda = 0$. This, together
with the gauge fixing condition, implies that the integration over $\Sca_\mu$
is equivalent to setting $\Sca_\mu = 0$ everywhere in the path integral,
which reduces eq.~\pref{ExtLagr} to \pref{LELWLagr}, as claimed. 

The dual version of the lagrangian is found by instead performing the
functional integrals in a different order, integrating out $\phi$ and
$\Sca_\mu$ and leaving $b_\mu$ as the dual variable. Care must be taken
when performing these integrals to properly handle the vortex boundary
condition which is satisfied by $\phi$, namely:
\eq
\label{vortexbc}
\phi(\vartheta + 2\pi) = \phi(\vartheta) + 2 \pi \, q_\phi \sum_a N_a ,
\eeq
where $N_a$ are integers labelling the vorticity of each vortex, and 
$\vartheta$ is the angular polar coordinate taken at spatial infinity,
a long distance away from the vortex positions. 

To integrate over $\phi$ it is convenient to write $\phi = \omega
+ \varphi$, where $\omega$ is a particular configuration having the
same boundary condition as does $\phi$, so $\varphi$ is simply
periodic: $\varphi(\vartheta + 2\pi) = \varphi(\vartheta)$. For
$\omega$ we choose
\eq
\label{omegadef}
\omega(x) = {2\pi \over q_\phi} \sum_a N_a \arctan\left({x^1 - y^1_a
\over x^2 - y^2_a} \right) ,
\eeq
where $y^i_a, i=1,2$ are the coordinates of the positions of each
vortex, with the index $i$ labelling the two space directions. 

Notice that the gauge potential defined by $v_\mu = \partial_\mu \omega$,
has vanishing field strength, {\it except} at the positions of the vortices,
where it has $\delta$-function singularities, so:
\eq
\label{omegafs}
\epsilon^{\mu\nu\lambda} b_\mu \partial_\nu v_\lambda =
- {2\pi \over q_\phi} \sum_a N_a y^\mu_a(t) b_\mu \delta[x - y_a(t)] .
\eeq
This term appears in the dual lagrangian, and provides the minimal coupling
of the vortex positions to the potential $b_\mu$.

With these definitions, the integrations over $\varphi$ and
$\Sca_\mu$ are straightforward. It is simplest to choose $\varphi = 0$ as a gauge
condition, and then directly perform the unconstrained Gaussian integral over
$\Sca_\mu$. The result is eq.~\pref{dualLagr}, without the vortex kinetic
term.

\bigskip
\begin{center}
{\bf APPENDIX D\\
Particle-Vortex Duality and Landau Level Addition}
\end{center}

In this appendix we derive the duality transformation
between pseudo-particles and vortices, but within a second-quantized 
path-integral framework. (See also ref.~\cite{manu} for a more detailed
discussion of particle/vortex physics within the abelian Higgs model
supplemented by a Chern Simons term.)  The treatment here is based
on \cite{Karlhede}, except that we use a relativistic notation
since this makes the manipulations simpler and, as shown in Appendix B, the
non-relativistic form is easily recovered from the relativistic form.

We start from the action for a complex scalar field $\Phi$,
with charge $e$, coupled to
an external electromagnetic field $A_\mu$
with a statistical gauge field $a_\mu$ 
\begin{equation}
\label{atildeaction}
S[\Phi,A,a] = -\int d^3x\left[
\frac{\pi e^2}{2\theta h}\epsilon^{\mu\nu\rho} a_\mu\partial_\nu a_\rho \right] + S_m[\Phi,\tilde a] 
\label{action1}
\end{equation}
with
\begin{eqnarray}
\label{eqthis}
&&S_m[\Phi,\tilde a] =\nonumber\\
&&-\frac{1}{2}\int d^3x \left[(i\hbar\partial_\mu-e\tilde a_\mu)\Phi\right]^\dagger
\left[(i\hbar\partial^\mu
-e\tilde a^\mu)\Phi \right] + S_{int}[|\Phi|^2]
\end{eqnarray}
where $\tilde a=A+a$, and $S_{int}$ is an interaction term, possibly including a mass term. 

\topic{The Flux Attachment Transformation}

The effective action involving the statistical gauge field
should reproduce the same physics as (\ref{origactionP}), if $\theta=2k\pi$.
To examine this further we first observe that the statistical gauge field
$a$ only ever appears in the action (\ref{eqthis}) 
in the combination $A+a$, so define $\tilde a = \eta A+a$ ,
(at the moment $\eta=1$, but it is introduced here for later convenience).  
Now integrating the matter fields out of (\ref{ZAa}) gives rise
to an effective action for $A_\mu$ and $a_\mu$ of the form 
\begin{eqnarray}
\label{stataction}
\Gamma^{(a)}_{eff}[A,a]&=&-\int d^3\bar p\left\{
{1\over 2}\Pi_1(p)
\tilde a_\mu(-p)\Lambda^{\mu\nu}\tilde a_\nu(p)\right.\nonumber\\
&&\qquad\qquad\left.+{1\over 2}\Pi_3(p)J^{\mu\nu}\tilde a_\mu(-p)
\tilde a_\nu(p)
\right\}\nonumber\\
&& -\left({e^2\pi\over 2h\theta}\right)\int d^3\bar p
\frac{\sqrt{p^2}}{\hbar}J^{\mu\nu}a_\mu(-p)a_\nu(p).
\end{eqnarray}
Even though the matter integrations cannot be done explicitly, the
construction ensures that the form factors $\Pi_1$ and $\Pi_3$
appearing in equation (\ref{stataction}) must be identical to those
appearing in (\ref{origactionP}). They will be functions of the
field strengths for $\tilde a_\mu$ in general but, provided $\tilde a_\mu$
is small ({\it i.e.} provided $a_\mu$ almost exactly cancels the
external field $\eta A_\mu$), they can be evaluated at $\tilde a_\mu=0$.

Now equation
(\ref{stataction}) should describe exactly the same physics as 
(\ref{origactionP}), after $a$ is integrated out, and this is what gives
rise to the flux attachment transformation, as we now describe. 

Integrating the statistical gauge field out of (\ref{stataction})
is easily achieved when $\tilde a_\mu$ is set to zero in $\Pi_1$ and $\Pi_3$, 
as it is then quadratic in $a_\mu$. The resulting effective action for $A_\mu$
will have the same form as (\ref{origactionP}) with $\Pi_1$ and $\Pi_3$ 
evaluated at $A_\mu=0$, but with different form factors, which we shall
denote by ${\tilde \Pi}_1$ and $\tilde{\Pi}_3$.

A subtlety in the integration is that in the quadratic from 
$a_\mu M^{\mu\nu}a_\nu$ the matrix $M^{\mu\nu}$ is not invertible as it has
a zero eigenvalue. However it is only really necessary to find a matrix
$\cal M$ such that, in relativistic formalism in momentum space, 
${\cal M}_{\mu\nu}M^{\nu\rho}={\delta_\mu}^\rho-{p_\mu p^\rho\over p^2}$
in order to integrate out $a$.
The result is \cite{KLZ} 
\begin{eqnarray}
\tilde \Gamma_{eff}[A]&=&-\int d^3\bar p\left\{
{1\over 2}\tilde\Pi_1(p)
A_\mu(-p)\Lambda^{\mu\nu}A_\nu(p)\right.\nonumber\\
&&\qquad\qquad\left.+{1\over 2}\tilde\Pi_3(p)J^{\mu\nu}A_\mu(-p)
A_\nu(p)
\right\},
\end{eqnarray}
where
\begin{eqnarray}
\label{Pi}
\tilde\Pi_1&=&\left({e^2\eta\pi\over h\theta}\right)^2{p^2\Pi_1\over \hbar^2D}
\nonumber\\
\tilde\Pi_3&=&\frac{\sqrt{p^2}}{\hbar}\left({e^2\eta^2\pi\over h\theta}\right)
\nonumber\\
&&\qquad
-\left(\frac{p^2}{\hbar^2}\right)\left({e^2\eta\pi\over h\theta}\right)^2
\left({\Pi_3 +\frac{\sqrt{p^2}}{\hbar}({e^2\pi\over h\theta})\over D}\right),
\end{eqnarray}
with
\begin{eqnarray}
D&=&\Pi_1^2 +\left[\Pi_3
+\frac{\sqrt{p^2}}{\hbar}\left({e^2\pi\over h\theta}\right)\right]^2\nonumber\\
&=&\frac{p^2}{\hbar}\left[\frac{p^2}{\hbar}\pi_1^2+\left(\pi_3 +
{e^2\pi\over h\theta}\right)^2\right].
\end{eqnarray}

The non-relativistic transformation is obtained by re-introducing the 
metric (\ref{metric}) and noting that
$p^2=g^{\mu\nu}p_\mu p_\nu=-p_0^2 +{\pi_2\over\pi_1}{\bf p}^2$,
and then equations ({\ref{Pi}) become
\begin{eqnarray}
\label{pi}
\tilde\pi_1&=&\left({e^2\eta\pi\over h\theta}\right)^2{\pi_1\over d}\nonumber\\
\tilde\pi_2&=&\left({e^2\eta\pi\over h\theta}\right)^2{\pi_2\over d}\nonumber\\
\tilde\pi_3&=&\left({e^2\eta^2\pi\over h\theta}\right)
-\left({e^2\eta\pi\over h\theta}\right)^2
\left({\pi_3 +({e^2\pi\over h\theta})\over d}\right),
\end{eqnarray}
where $d=(-\pi_1^2p_0^2 + \pi_1\pi_2{\bf p}^2)/\hbar^2 + 
\left(\pi_3 + {e^2\pi\over h\theta}\right)^2$, and this is
the form of the transformation given in \cite{KLZ} (except it
is given in Euclidean signature in that reference).

For a Hall conductor the transverse conductivity
is related to the polarization tensor $\Pi_3$ by 
\begin{equation}
\sigma_{xy}=\pi_3=\hbar\Pi_3/\sqrt{p^2}
\end{equation}
while the transverse conductivity
involves breaking the 3-momentum
up into frequency and spatial momentum $p_\mu=(\hbar\omega,{\bf p})$
and taking the limit
\begin{eqnarray}
\label{conductivity}
\sigma_{xx}&=&
\lim_{\omega\rightarrow 0}\left[\sqrt{p^2}\pi_1/\hbar\right]_{{\bf p}=0}
=\lim_{\omega\rightarrow 0}\left[\hbar\Pi_1/\sqrt{p^2}\right]_{{\bf
p}=0}\nonumber\\
&=&-i\lim_{\omega\rightarrow 0}\left[{\Pi_1/\omega}\right]_{{\bf p}=0}.
\end{eqnarray} 
The transformed conductivities, $\tilde\sigma_{xx}$ and 
$\tilde\sigma_{xy}$ are related to the transformed polarization tensors
$\tilde\Pi_1$ and $\tilde\Pi_3$ in a similar way.

Equation (\ref{pi}) then gives,
with $\eta=1$, $\theta=2k\pi$ and units in which $e^2/h=1$,
\begin{eqnarray}
\label{condmap}
\tilde\sigma_{xx}&=&{\sigma_{xx}\over {4k^2(\sigma^2_{xx}+\sigma_{xy}^2)
-4k\sigma_{xy}+1}}\nonumber\\
\tilde\sigma_{xy}&=&{2k(\sigma_{xx}^2+\sigma_{xy}^2)+\sigma_{xy}\over 
{4k^2(\sigma^2_{xx}+\sigma_{xy}^2) -4k\sigma_{xy}+1}}.
\end{eqnarray}
It is stressed here that the above analysis represents a 
symmetry under certain conditions, as discussed in \cite{KLZ},
such as very low temperatures.
In the long-wavelength, zero frequency limit, 
the phase diagram of the quantum Hall effect is symmetric at low temperatures under the
above transformation.
For example the critical exponents
at related second order phase transitions should be identical.

It should be remembered that $\sigma_{xx}$ and $\sigma_{xy}$ represent the components of a tensor, it may at first sight seem unnatural
to be applying a non-linear map which mixes up the different components
of a tensor --- what about co-variance of the tensor components?
In fact equations \pref{condmap} are very natural from this point of view.
If we define a complex co-ordinate $z:=x+iy$, and it's conjugate
$\bar z=x-iy$, then
the conductivity tensor in these co-ordinates is reduced to
a single quantity, $\sigma:=\sigma_{xy}+i\sigma_{xx}$, with
a positive imaginary part, since $\sigma_{xx}>0$.
The transformation reduces, in this co-ordinate system, 
to 
\begin{equation}
\label{etasigma}
\tilde\sigma=\eta^2{\sigma\over\left(1+{\theta\over\pi}\sigma\right)},
\end{equation}
which gives
$\tilde\sigma = {\sigma\over 1+2k\sigma}$ for $\eta=1$ and $\theta=2k\pi$.
This last form can be obtained by $k$ iterations of the generating
transformation 
\begin{equation}
\label{ST2S}
\tilde\sigma = {\sigma\over 1+2\sigma},
\end{equation}
as is easily checked. Equation (\ref{ST2S}) is the transformation $ST^{-2}S$
in the text.

\topic{Particle-Vortex Duality}

We could consider eq.~\pref{eqthis} as either: 
i) a bosonic problem which is transformed from
another bosonic problem ($\theta = 2k\pi$) or ii) a bosonic problem
transformed from a fermionic problem ($\theta = (2k+1)\pi$) ---
only in the former case is the statistical gauge field transformation
a symmetry. 

It will be argued in this section, following \cite{Karlhede}, that there is a second
symmetry in the bosonic case ($k$ even), $\sigma\rightarrow -1/\sigma$.
This is a ${\bf Z}_2$ symmetry which maps an insulator $\sigma=0$   to
a superconductor $\sigma=i\infty$ and 
has $\sigma_{xx}=e^2/h$, $\sigma_{xy}=0$ as a fixed point.

We derive the second duality transformation as follows:
firstly write (\ref{atildeaction}) in terms of the pseudo-particle
para-magnetic current,
$j_\mu=\frac{ie\hbar}{2}[\Phi^\dagger\partial_\mu\Phi-
(\partial_\mu\Phi)^\dagger\Phi]$, as
\begin{eqnarray}
\label{aaction1}
S[\Phi,A,a]&=&-\int d^3x\left[
\frac{\pi e^2}{2\theta h}\epsilon^{\mu\nu\rho} 
a_\mu\partial_\nu a_\rho \right] \nonumber\\
&&\kern -1cm +\int d^3x \Bigl[-\frac{\hbar^2}{2}(\partial_\mu\Phi^\dagger)
(\partial^\mu\Phi) - \frac{e^2}{2}\vert\Phi\vert^2\tilde a^\mu\tilde a_\mu
+\tilde a^\mu j_\mu \Bigr]\nonumber \\ 
&+&S_{int}[\vert\Phi\vert^2].
\end{eqnarray}

Alternatively the action (\ref{action1}) can be written in terms of the vortex
current by splitting $\Phi$ up into a smooth part and a vortex
part as 
\begin{equation} 
\Phi({\bf r}) = \Phi_0({\bf r}) e^{-i\vartheta({\bf r})
}v({\bf r})  
\end{equation} 
where $\Phi_0({\bf r}) $ is real,
$\vartheta({\bf r}) $ is
real, positive and single valued and 
\begin{equation} 
v({\bf r})  =
e^{i{2\pi\over q_\phi}\sum_a N_a\arctan\left(\frac{x^1-y_a^1}{x^2-y_a^2}\right)}, 
\end{equation} 
where $(y^1_a,y^2_a)$ denote the position of the vortex labelled by $a$, 
which should be summed over in the
path integral.
In these variables the matter action reads
\begin{eqnarray}
\label{Bosaction}
S_m[\Phi_0,\vartheta,\tilde a] &=& 
 -\frac{1}{2}\int d^3x \Bigl[(\hbar\partial_\mu\Phi_0)^2\nonumber\\
&+&\Phi_0^2(\hbar\partial_\mu\vartheta + i\hbar v^*\partial_\mu v -e\tilde a_\mu)^2
\Bigr]
+S_{int}[|\Phi|^2],
\end{eqnarray}
where indices are understood to be contracted with the Minkowski metric $(-1,+1,+1)$.

We can now perform the integral over $\vartheta$ by introducing
an auxiliary vector field, $\lambda_\mu=\hbar\partial_\mu\vartheta$, and imposing the constraint
$\partial_\mu\lambda_\nu-\partial_\nu\lambda_\mu=0$ with a Lagrange
multiplier field, ${ \tilde b}_\mu$. So we write
\begin{eqnarray}
&&\int{\cal D}\vartheta e^{-\frac{i}{2\hbar}\int d^3 x 
\Phi_0^2(\hbar\partial_\mu\vartheta + i\hbar v^*\partial_\mu v -e\tilde a_\mu)^2}
=\nonumber\\
&&\int{\cal D}\lambda{\cal D}b  e^{-\frac{i}{2\hbar}\int d^3 x 
\Phi_0^2(\lambda_\mu + i\hbar v^*\partial_\mu v -e\tilde a_\mu)^2
-\frac{i}{\hbar e} 
\int d^3 x \epsilon^{\mu\nu\rho}{\tilde b}_\mu\partial_\nu \lambda_\rho}.
\end{eqnarray}

Performing the functional integral over $\lambda_\mu$ puts (\ref{action1})
in the form
\begin{eqnarray}
\label{abaction}
S&[&\Phi_0,A,a,b]=\int d^3x\Bigl[
-\frac{\pi}{2\theta}\frac{e^2}{h}\epsilon^{\mu\nu\rho} a_\mu\partial_\nu a_\rho \nonumber\\
&&-\epsilon^{\mu\nu\rho}\tilde a_\mu \partial_\nu {\tilde b}_\rho 
+{\tilde j}^\mu{\tilde b}_\mu 
-\frac{1}{4 e^2\Phi^2_0}{\tilde f}^b_{\mu\nu}({\tilde f}^b)^{\mu\nu} \nonumber\\
&&\qquad-\frac{1}{2}\hbar^2\partial_\mu\Phi_0\partial^\mu\Phi_0\Bigr]
+S_{int}^\prime[\Phi_0^2],
\end{eqnarray}
where ${\tilde f}^b_{\mu\nu}=\partial_\mu {\tilde b}_\nu-\partial_\nu {\tilde b}_\mu$ is the
field strength for ${\tilde b}_\mu$
and the vortex current is 
$\tilde{j}^\mu=\frac{i\hbar}{e}\epsilon^{\mu\nu\lambda}(\partial_\nu v^*)(\partial_\lambda v)$.
The integration over $\lambda$ also induces a 
functional determinant, $\ln(\hbox{det}\Phi_0)$ and this has been absorbed
into the interaction term for $\Phi_0$, as indicated by the prime
on $S_{int}^\prime[\Phi_0^2]$.

Now integrate $a_\mu$ out of (\ref{abaction}).
The only terms involving $a$ are
\begin{equation}
S_a = -\int d^3x\left[
\frac{\pi}{2\theta}\frac{e^2}{h}
\epsilon^{\mu\nu\lambda}a_\mu\partial_\nu a_\lambda +
\epsilon^{\mu\nu\lambda}a_\mu\partial_\nu {\tilde b}_\lambda\right].
\end{equation}
So integrating out $a$ we get a term
\begin{equation}
\int d^3x \left[ \frac{\theta}{2\pi}\frac{h^2}{e}\epsilon^{\mu\nu\lambda}{\tilde b}_\mu
\partial_\nu {\tilde b}_\lambda
\right] .
\end{equation}
The action in terms of ${\tilde b}$ is now
\begin{eqnarray}
S^{(b)}&[&\Phi_0,A,{\tilde b}] = \int d^3x\left[
\frac{\theta}{2\pi}\frac{h}{e^2}\epsilon^{\mu\nu\rho} {\tilde b}_\mu\partial_\nu 
{\tilde b}_\rho\right]\nonumber\\
&& +\int d^3x\left[-\frac{1}{4 e^2\Phi_0^2}{\tilde f}^b_{\mu\nu}({\tilde f}^b)^{\mu\nu} 
- \epsilon^{\mu\nu\lambda}A_\mu\partial_\nu {\tilde b}_\lambda
+{\tilde b}^\mu\tilde{j}_\mu \right] \nonumber \\
&&\qquad- \int d^3x\left[\frac{\hbar^2}{2}\partial_\mu\Phi_0\partial^\mu\Phi_0   \right]
+S_{int}^\prime[\Phi^2_0].
\end{eqnarray}
Now let
${\tilde b}= b +\frac{e^2}{h}\frac{\pi}{\theta}A$ 
and the action becomes
\begin{eqnarray}
\label{noVb}
S^{(b)}[\Phi_0,A,b]&=&-\int d^3x\left[
\frac{\pi}{2\theta}\frac{e^2}{h}
\epsilon^{\mu\nu\rho}A_\mu\partial_\nu A_\rho\right]\nonumber\\
&-&\int d^3x\left[\frac{\pi}{2\tilde\theta}\frac{h}{e^2}
\epsilon^{\mu\nu\rho} b_\mu\partial_\nu b_\rho
\right] \nonumber \\
&+& \int d^3x\left[-\frac{1}{4e^2\Phi_0^2}\tilde f^b_{\mu\nu}(\tilde f^b)^{\mu\nu} 
+\tilde{b}_\mu\tilde{j}^\mu \right]\nonumber\\ 
&-& \int d^3x\left[\frac{\hbar^2}{2}\partial_\mu\Phi_0\partial^\mu\Phi_0   \right]
+S_{int}^\prime[\Phi^2_0],
\end{eqnarray}
where 
$\tilde\theta=-\frac{\pi^2}{\theta}$.

For comparison (\ref{aaction1}) reads, 
writing $\Phi=\Phi_0e^{-i\vartheta}$ so that the pseudo-particle
para-magnetic current is $j_\mu=e\hbar\Phi_0^2\partial_\mu\vartheta$,
\begin{eqnarray}
\label{noVa}
S^{(a)}[\Phi_0,A,a]&=&-\int d^3x\Bigl[
\frac{\pi}{2\theta}\frac{e^2}{h}
\epsilon^{\mu\nu\rho} a_\mu\partial_\nu a_\rho\Bigr]\nonumber\\
&&+\int d^3x\Bigl[-\frac{e^2\Phi_0^2}{2}\tilde a_\mu \tilde a^\mu+\tilde a^\mu j_\mu \Bigl]\nonumber\\
&&\kern -25pt -\int d^3x \Bigr[\frac{\hbar^2}{2}(\partial_\mu\Phi_0)^2
+\frac{1}{2e^2\Phi_0^2}j_\mu j^\mu\Bigr] 
+S_{int}[\Phi_0^2].
\end{eqnarray}

The duality symmetry that we seek lies in the
symmetry between (\ref{noVb}) and (\ref{noVa}).
One way to see this is to fix the external field $A_\mu$ at some background
value and treat $\Phi_0$ as a classical field.
If $\Phi_0$ is non-zero constant the action (\ref{noVa}) describes
a superconductor while (\ref{noVb}) describes
an insulator --- as argued in \cite{Fisher} vortices
condense to give an insulator.
If the gauge symmetry is not broken $S^{(a)}$ and $S^{(b)}$
describe conductors dual to one another. To examine 
this situation we analytically continue
$\Phi_0^2$ to complex values and set
$e^2\Phi_0^2\approx \frac{e^2}{h}\frac{\sqrt{p^2}}{\hbar}C=\frac{ie^2}{h}\omega C$
where $C$ is dimensionless (this is the leading term
in the Fourier expansion of $\Phi_0^2$, in the long wavelength 
limit when the gauge symmetry
is not broken). So the longitudinal conductivity is (setting 
$p^\mu=(\hbar\omega,{\bf 0})$)
\begin{equation} 
\sigma_{xx}^a=\hbar e^2\Phi_0^2/\sqrt{p^2}=
\lim_{\omega\rightarrow 0}\left(\frac{e^2\Phi_0^2}{i\omega}\right)=C\frac{e^2}{h}
\end{equation} 
(see equation (\ref{conductivity}) with $\pi_1=\hbar^2e^2\Phi_0^2/p^2$).
Similarly $S^{(b)}$ describes a conductor 
with longitudinal conductivity
\begin{equation} 
\sigma_{xx}^b=\lim_{\omega\rightarrow 0}\left(\frac{i\omega}{e^2\Phi_0^2}\right)=
\frac{1}{C}\frac{h}{e^2},
\end{equation}
(equation (\ref{conductivity}) with $\pi_1=1/(e\Phi_0)^2$).

For a fixed external field $\Phi_0$, and so the 
conductivities $\sigma^a_{xx}$ and $\sigma^b_{xx}$,
depend on the external field as well as other 
external parameters, such as the temperature. In general, therefore, it
should be possible to find pairs of values for the external 
parameters, labelled generically by $X$, such that
$C(X)=1/C(X^\prime)$. (this is equivalent to setting ${\bf \tilde V}={\bf P}$
in section IV). Then the effective actions \pref{noVb} and \pref{noVa} are identical
except for the extra Chern-Simons term for $A_\mu$ in \pref{noVb}.

In particular if we take $\theta=-\pi$ in (\ref{noVb})
and $\pi$ in (\ref{noVa}) (these two values of $\theta$ are of course indistinguishable
in the statistical gauge field transformation) the sole effect of the extra term
on the effective action is to shift the transverse conductivity by 
$\sigma_{xy}\rightarrow\sigma_{xy}+1$ (where we have set $e^2/h=1$). Of course 
$\theta=\pm\pi$ means that
we started from fermions and transformed to bosons using the statistical gauge field so
we can argue that we have derived the $T$ transformation (which corresponds to
Landau-Level addition in the first quantised theory) as follows:
start with a fermionic system and transform to a bosonic system using $\theta=\pm\pi$,
then use the above argument to show that $\sigma\rightarrow\sigma+1$ is a symmetry
and then transform back to fermions. This shows that $\sigma\rightarrow\sigma+1$
is a symmetry for the fermionic system (it is not a symmetry for a
bosonic system, because $\theta=\pm\pi$ does not keep bosons as bosons).
This argument, while plausible,
has ignored the currents and the functional integration over $\Phi_0$.

A more convincing argument takes the currents, both pseudo-particle and
vortex, into account.
Pseudo-particles and vortices are massive and so should decouple in the long wavelength limit.
Just as in the
derivation of the flux attachment transformation we can argue that, in the
long wavelength limit, integrating out $\Phi_0$ from 
(\ref{noVa}) and summing over pseudo-particle currents must lead to
an effective action for the external gauge field of the form
\begin{eqnarray}
\label{aaction}
{\Gamma^{(a)}_{eff}}&[&A,a]=\nonumber\\
&&-\int d^3xd^3x^\prime\left\{
\left({1\over 4}\right) (\tilde{f}^a)_{\mu\nu}(x)\pi^a_1(x-x^\prime)
(\tilde{f}^a)^{\mu\nu}(x^\prime)\right.\nonumber\\
&&\left.+{1\over 2}\epsilon^{\mu\nu\lambda}\tilde a_\mu(x)\pi^a_3(x-x^\prime)
\partial_\nu\tilde a_\lambda(x^\prime)\right\}\nonumber \\
 &&\qquad -\frac{e^2}{h}\left({\pi\over 2\theta}\right)\int d^3x\epsilon^{\mu\nu\lambda}
a_\mu(x)\partial_\nu a_\lambda(x),
\end{eqnarray}
with $\tilde f^a_{\mu\nu}=\partial_\mu\tilde a_\nu - \partial_\nu\tilde a_\mu
=\partial_\mu(a_\nu + A_\nu)-\partial_\nu(a_\mu + A_\mu)$
(we have set $\pi^a_2=\pi^a_1$ in the relativistic form --- the superscript $a$
indicates that these are polarization tensors associated with the field $a_\mu$).
As before integrating out $\tilde a$ now gives an effective action for the
external field $A$,
\begin{eqnarray}
\label{tildeaction}
\tilde \Gamma_{eff}[A]&=&-{1\over 2}\int d^3\bar p\left\{
{p^2\over\hbar^2}\tilde\pi_1(p)
A_\mu(-p)\Lambda^{\mu\nu}A_\nu(p)\right.\nonumber\\
&&\qquad\qquad\left.+{\sqrt{p^2}\over\hbar}\tilde\pi_3(p)J^{\mu\nu}A_\mu(-p)
A_\nu(p)
\right\},
\end{eqnarray}
with polarizations
\begin{eqnarray}
\label{Pia}
\tilde\pi_1(\omega,{\bf p})&=&\left({e^2\pi\over h\theta}\right)^2{\pi^a_1\over d^a}\nonumber\\
\tilde\pi_3(\omega,{\bf p})&=&{e^2\pi\over h\theta}
-\left({e^2\pi\over h\theta}\right)^2\left({\pi^a_3+({e^2\pi\over h\theta}) \over d^a}\right),
\end{eqnarray}
where $d^a=(\pi^a_1)^2(-p_0^2+{\bf p}^2)/\hbar^2+(\pi^a_3 +({e^2\pi\over h\theta}))^2$.
This leads to complex conductivities that are related by 
\begin{equation}
\label{sigmat}
\tilde\sigma = {\sigma^a\over (1+{\theta\over\pi}\sigma^a )},
\end{equation}
with $\sigma^a=\lim_{\omega\rightarrow 0}(\pi^a_3-\omega\pi^a_1)$ and $e^2/h=1$
(remember $\lim_{\omega\rightarrow 0}(\omega\pi^a_1)=-i\sigma^a_{xx}$).

Similarly integrating out $\Phi_0$ and summing over vortex configurations
in (\ref{noVb}) must lead, in the
long wavelength limit, to
\begin{eqnarray}
\label{baction}
\Gamma^{(b)}_{eff}&[&A,b]=\nonumber\\
&&-\int d^3xd^3x^\prime\left\{
\left({1\over 4}\right) (\tilde{f}^b)_{\mu\nu}(x)\pi^b_1(x-x^\prime)(\tilde
{f}^b)^{\mu\nu}(x^\prime)\right.\nonumber\\
&&\left.+{1\over 2}\epsilon^{\mu\nu\lambda}\tilde b_\mu(x)\pi^b_3(x-x^\prime)
\partial_\nu\tilde b_\lambda(x^\prime) 
\right\}\nonumber\\
&&-\frac{e^2}{h}\int d^3x\frac{\pi}{2\theta}\epsilon^{\mu\nu\rho}A_\mu\partial_\nu A_\rho
\nonumber \\
&& \qquad -{\pi\over2\tilde\theta }\frac{h}{e^2}\int d^3x\epsilon^{\mu\nu\lambda}b_\mu(x)
\partial_\nu b_\lambda(x).
\end{eqnarray}
Remember that here
$(\tilde{f}^b)_{\mu\nu}=
\partial_\mu(b_\nu+{e^2\over h}{\pi\over\theta}A_\nu)-
\partial_\nu(b_\mu+{e^2\over h}{\pi\over\theta}A_\mu)$.

Now we can integrate $b$ out of (\ref{baction}) in exactly the same way as
$a$ was integrated out of (\ref{aaction}) ---  the only differences are 
i) the presence of the  term involving 
$-\frac{e^2}{h}\frac{\pi}{2\theta}\epsilon^{\mu\nu\rho}A_\mu\partial_\nu A_\rho$ which
subtracts ${e^2\pi\over h\theta}$ from the right hand side of 
$\tilde \pi_3$ in (\ref{Pia}),
ii) $\frac{e^2\pi}{h\theta}$ is replaced with $\frac{h\pi}{e^2\tilde\theta}$
with $\tilde\theta=-{\pi^2\over \theta}$
in (\ref{Pia});
iii) we should use $\eta={e^2\pi\over h\theta}$
and iv) the $\pi^a_i$ are replaced by $\pi^b_i$, which will in general be different. 
The resulting effective action is
\begin{eqnarray}
\label{Gammaeff}
\tilde{\tilde \Gamma}_{eff}[A]&=&-{1\over 2}\int d^3\bar p\left\{
{p^2\over\hbar^2}\tilde{\tilde\pi}_1(p)
A_\mu(-p)\Lambda^{\mu\nu}A_\nu(p)\right.\nonumber\\
&&\qquad\qquad\left.+{\sqrt{p^2}\over\hbar}\tilde{\tilde\pi}_3(p)J^{\mu\nu}A_\mu(-p)
A_\nu(p)
\right\},
\end{eqnarray}
with different polarization transformations to (\ref{Pia}) --- see equation
(\ref{pi}),
\begin{eqnarray}
\label{Pib}
\tilde{\tilde\pi}_1(\omega,{\bf p})&=&
\left({\pi\over\theta}\right)^2
\left({\pi\over\tilde\theta}\right)^2{\pi^b_1\over d^b}={\pi^b_1\over d^b}\nonumber\\
\tilde{\tilde\pi}_3(\omega,{\bf p})&=&
\left({e^2\pi\over h\theta}\right)^2{\pi\over\tilde\theta}
-\left({\pi\over\theta}\right)^2\left({\pi\over\tilde\theta}\right)^2
\left({\pi^b_3 +({h\pi\over e^2\tilde\theta})\over d^b}\right)
+{e^2\pi\over h\theta}\nonumber\\
&=&{\left(({h\theta\over e^2\pi})-\pi^b_3\right) \over d^b},
\end{eqnarray}
where $d^b=(\pi^b_1)^2(-p_0^2+{\bf p}^2)/\hbar^2+\bigl(\pi^b_3 -(\frac{h\theta}{e^2\pi})\bigr)^2$.
 
 In terms of complex conductivities (\ref{Pib}) results in (again with $e^2/h=1$)
\begin{equation}
\label{sigmatt}
\tilde{\tilde\sigma}=-{1\over (\sigma^b-{\theta\over\pi})},
\end{equation}
with $\sigma^b=\lim_{\omega\rightarrow 0}(\pi^b_3-\omega\pi^b_1)$.
If $\theta=2\pi k$ then this equation represents a symmetry of a bosonic theory,
$\sigma\rightarrow-\frac{1}{\sigma - 2k}$ which is equivalent to the modular
transformation 
$\gamma=
\left(
\matrix{0&1\cr -1&2k\cr}
\right)$.  The particular case $k=0$ is $\sigma\rightarrow -1/\sigma$.

As discussed in Appendix B the effective actions \pref{tildeaction} and (\ref{Gammaeff}) are 
long-wavelength but not necessarily weak field if the polarisations
depend on the background electric and magnetic fields. 
This means that the conductivities $\sigma^a$ and $\sigma^b$ will
depend on external parameters, generically labelled $X$ before, including the
external magnetic field and the temperature, so consider (\ref{Gammaeff}) to
be the effective action for a perturbation on the external field. 
Typically it will be possible to
find pairs of parameters so that $\sigma^a(X)=\sigma^b(X^\prime)$.
Then eliminating $\sigma^a$ and $\sigma^b$ from
(\ref{sigmat}) and (\ref{sigmatt}) 
gives $S$ ($\tilde{\tilde\sigma}=-1/\tilde\sigma$) 
when $\theta=0$. Together with the flux attachment
transformation \pref{ST2S}
this generates the group $\Gamma_\theta$ for bosons.
Conjugating with $ST$ as described in the text gives $\Gamma_0(2)$ for fermions.

\begin{center}
{\bf APPENDIX E:\\
The Gaussian Integrals}
\end{center}

In this appendix we perform in detail the Gaussian integrations which lead to
our central formula, eq.~\pref{PiPV}. Our starting point is eq.~\pref{dualLagr}
after the integration over the positions of the quasi-particles and vortices has been
performed, using eqs.~\pref{RFdef2a} and \pref{RFdef2b}. This
starting point may be written:
\eqa
S_0 &=& -\int d^3x \left[ {\pi \over 2 \theta} \epsilon^{\mu\lambda\nu} a_\mu
\partial_\lambda a_\nu +\epsilon^{\mu\lambda\nu} b_\mu \partial_\lambda
(a+A)_\nu \right] \nonumber\\
&&-\frac12 \int d^3x\, d^3y \; \Bigl[ (a+A)_\mu(x) P^{\mu\nu}(x-y) (a+A)_\nu(y)\nonumber\\
&&\qquad \qquad + b_\mu(x) V^{\mu\nu}(x-y) b_\nu(y) \Bigr]. 
\eeqa

Our task is to perform the integrals over $a_\mu$ and $b_\mu$, and our interest
is in the dependence of the result on $A_\mu$. Because none of the functional determinants
which arise in the integrations depend on $A_\mu$, these may be neglected as contributing
an $A_\mu$-independent additive constant to the EM response function, $\Gamma[A]$. 

Performing first the $b_\mu$ integration, the saddle-point condition may be written as:
\eq
b_\mu(x) = -\int d^3y \; {\cal V}_{\mu\nu}(x-y) \epsilon^{\nu\alpha\beta}
\partial_\alpha (a+A)_\beta(y),
\eeq
where the kernel ${\cal V}_{\mu\nu}$ satisfies the definition
\eq
\int d^3y \; {\cal V}_{\mu\lambda}(x-y) V^{\lambda\nu}(y-z) = {\Lambda_\mu}^\nu(x-z),
\eeq
for the projection operator \hfill\break${\Lambda_\mu}^\nu(x-z) = (\delta_\mu^\nu - 
\partial_\mu \partial^\nu/\partial^2) \; \delta^3(x-z)$. 

The result for the effective action then becomes:
\eqa
S_1 &=& -\int d^3x \left[ {\pi \over 2 \theta} \epsilon^{\mu\lambda\nu} a_\mu
\partial_\lambda a_\nu \right] \\
&&-\frac12 \int d^3x\, d^3y \; (a+A)_\mu(x) \hat{P}^{\mu\nu}(x-y) (a+A)_\nu(y) , \nonumber 
\eeqa
where:
\eq
\hat{P}^{\mu\nu} = P^{\mu\nu}- \epsilon^{\mu\lambda\rho} 
\partial_\lambda {\cal V}_{\rho\alpha} \,
\epsilon^{\alpha\beta\nu}\partial_\beta .
\eeq

The construction of kernels like ${\cal V}_{\mu\nu}$ is dramatically simplified
by working in momentum space, and expanding the kernels in terms of the basis of tensors
${\Lambda_\mu}^\nu$ and ${J}^{\mu\nu}$, as introduced in the text. For instance,
if $V^{\mu\nu}(p) = A_1 \Lambda^{\mu\nu} + A_2 {J}^{\mu\nu}$, then
${\cal V}_{\mu\nu}(p) = B_1 \Lambda_{\mu\nu} + B_2 {J}_{\mu\nu}$
with
\eq
B_1 = {A_1 \over A_1^2 + A_2^2} , \qquad B_2 = - \; {A_2 \over
A_1^2 + A_2^2},
\eeq
which is more compactly written in terms of the complex variables ${\bf A}
= A_1 + i A_2$ and ${\bf B} = B_1 + i B_2$ as: ${\bf B} = 1/{\bf A}$.

Proceeding in the same vein, the saddle point for the $a_\mu$ integration
is
\eq
a_\mu = - \int d^3y \; {\cal K}_{\mu\nu} \hat{P}^{\nu\lambda} A_\lambda,
\eeq
where ${\cal K}_{\mu\lambda} K^{\lambda \nu} = {\Lambda_\mu}^\nu$ and
$K^{\mu\nu} = \hat{P}^{\mu\nu} + {\pi\over \theta} \, \epsilon^{\mu\lambda\nu}
\partial_\lambda$. Using this in the integration over $a_\mu$ gives the
electromagnetic response function of eq.~\pref{PIdef}, with the EM response
tensor, $\Pi^{\mu\nu}$ given by:
\eq
\Pi^{\mu\nu} = \hat{P}^{\mu\nu} - \hat{P}^{\mu\lambda} {\cal K}_{\lambda\rho}
\hat{P}^{\rho\nu}.
\eeq

Once expressed in terms of complex variables, as above, this result gives eq.~\pref{PiPV}
of the text.

\end{document}